\documentclass[fleqn,usenatbib]{mnras}

\usepackage{newtxtext,newtxmath}
\usepackage[capitalise]{cleveref}

\usepackage[T1]{fontenc}

\DeclareRobustCommand{\VAN}[3]{#2}
\let\VANthebibliography\thebibliography
\def\thebibliography{\DeclareRobustCommand{\VAN}[3]{##3}\VANthebibliography}

\def\ergs{{\rm erg}\,{\rm s}^{-1}}
\def\msun{{\rm M}_{\odot}}

\def\cMpc{{\rm cMpc}}
\def\cGpc{{\rm cGpc}}

\usepackage{graphicx}	
\usepackage{amsmath}	
\usepackage{comment}

\title[Quasars in FLAMINGO]{The luminosity function and clustering of bright quasars in the FLAMINGO cosmological simulations}

\author[B. Ding et al.]{
Boyi Ding,$^{1,4}$\thanks{E-mail: ding@strw.leidenuniv.nl}
Elia Pizzati,$^{1,5}$
Joop Schaye,$^{1}$
Joseph F. Hennawi$^{1,2}$,
William McDonald,$^{1}$
Matthieu Schaller$^{3,1}$
\\
$^{1}$Leiden Observatory, Leiden University, P.O. Box 9513, 2300 RA Leiden, The Netherlands\\
$^{2}$Department of Physics, University of California, Santa Barbara, CA 93106, USA\\
$^{3}$Lorentz Institute for Theoretical Physics, Leiden University, PO Box 9506, NL-2300 RA Leiden, The Netherlands \\
$^{4}$The Australian National University, Mount Stromlo Observatory, Cotter Road, Canberra, ACT 2611, Australia\\
$^{5}$Center for Astrophysics | Harvard \& Smithsonian, 60 Garden St., Cambridge, MA 02138, USA
}

\date{Accepted XXX. Received YYY; in original form ZZZ}

\pubyear{2015}

\begin{document}
\label{firstpage}
\pagerange{\pageref{firstpage}--\pageref{lastpage}}
\maketitle

\begin{abstract}
Cosmological hydrodynamical simulations are essential tools for studying the formation and evolution of galaxies and their central supermassive black holes. While they reproduce many key observed properties of galaxies, their limited volumes have hindered comprehensive studies of the AGN and quasar populations. In this work, we leverage the FLAMINGO simulation suite, focusing on its large $(2.8\,\mathrm{Gpc})^3$ volume, to investigate two key observables of quasar activity: the quasar luminosity function (QLF) and quasar clustering. FLAMINGO reproduces the observed QLF at low redshift ($z \lesssim 1$) and for faint quasars ($L_\mathrm{bol} \lesssim 10^{45}\,\mathrm{erg\,s^{-1}}$), but significantly underpredicts the abundance of bright quasars at $z \approx 1$-$3$. Introducing a 0.75 dex log-normal luminosity scatter to represent unresolved small-scale variability boosts the number of bright quasars by upscattering lower-luminosity systems, thereby improving agreement with observations at the bright end. A decomposition of the QLF by black hole mass reveals that this boost is primarily driven by low-mass black holes radiating above the Eddington limit. Nevertheless, limitations remain in fully reproducing the rise and decline of the bright quasar population over cosmic time and in matching the black hole masses inferred from quasar spectra. Thanks to FLAMINGO's large volume, we can robustly sample rare, luminous quasars and measure their spatial clustering for $\log_{10} L_\mathrm{bol}/\ergs \gtrsim 45.5$. The simulation reproduces the observed clustering across $0 \lesssim z \lesssim 3$, and the reduced luminosity dependence introduced by scatter aligns with observational trends. However, it underpredicts the clustering strength at $z \approx 4$, consistent with other models and possibly reflecting high-redshift observational uncertainties.
\end{abstract}




\section{Introduction} \label{sec:introduction}

    
Supermassive black holes (SMBHs) are believed to reside at the center of most galaxies \citep{Kormendy1995}. Observational evidence suggests a co-evolution between these black holes and their host galaxies, indicated by well-established correlations between black hole masses and the properties of their host galaxy spheroids, such as mass, luminosity, and velocity dispersion \citep{Kormendy1995, Ferrarese2000, Gebhardt2000, Gultekin2009}. Active Galactic Nuclei (AGN) and quasars, observable manifestations of accreting SMBHs, provide crucial insights into these processes. The accretion of matter onto SMBHs releases vast amounts of binding energy, resulting in high radiative efficiencies ($\epsilon_\mathrm{r}\sim 0.05-0.3$) that make quasars detectable at vast cosmic distances (up to $z\approx7.5$; see e.g., \citealt{wang2021, fan2023}).

Hydrodynamical cosmological simulations have emerged in the last few decades as powerful tools to study galaxy formation and evolution in a cosmological context. Landmark simulation projects such as Illustris \citep{vogelsberger2014}, EAGLE \citep{schaye2015}, and IllustrisTNG \citep{nelson2019} have shown that it is feasible to attain satisfactory agreements with observational constraints on galaxy properties over several Gyr of cosmic history, including galaxy stellar mass functions, galaxy star formation rates, galaxy clustering, morphologies, sizes, and colours. 

Critically, simulations have shown that the feedback mechanisms produced by quasars and AGN are essential to regulate the growth and evolution of galaxies, and to explain the interconnected growth of galaxies and SMBHs \citep[e.g.,][]{Di_Matteo_2008, Booth2009, Booth2011, Dubois_2015, Bower2017, Weinberger2018}. The amount of energy released by AGN feedback is proportional to the mass that
has been accreted onto the black holes, but the exact subgrid modelling of AGN feedback varies greatly in different simulations. For this reason, while modern simulations reliably produce populations of quiescent massive galaxies up to cosmic noon, they may do so with very different SMBH populations, and employing alternative prescriptions for SMBH seeding, growth, and evolution \citep[e.g.,][]{Habouzit2021}.  

Given the pivotal role played by quasars and AGN in modern large-scale cosmological simulations, assessing whether these populations are faithfully represented in simulations is an important question in the field. Several studies have shown that simulations are broadly capable of reproducing the growth and evolution of the black hole population, including the galaxy mass-BH mass relation, the black hole mass function, and the quasar/AGN luminosity function \citep{Degraf2012,Sijacki2015,RosasGuevara2016,Weinberger2018}. However, while the properties of simulated galaxies have been under extensive scrutiny in the last decades, the ability of current large-scale simulations to reliably model AGN and quasar properties remains much less clear, especially beyond the local Universe \citep{Habouzit2021, Habouzit2022}. In particular, most simulations calibrate black hole masses in order to reproduce the local black hole mass-galaxy mass relation, but do not employ any other observables related to quasar/AGN activity to constrain their subgrid models \citep[e.g.,][]{Springel2005,Di_Matteo_2008,Booth2009,Dubois_2012,vogelsberger2014}. For this reason, comparing such observables to real measurements constitutes a useful test of state-of-the-art cosmological simulations. 

What makes a comparison with quasar observables difficult in the context of cosmological simulations, however, is the remarkable rarity of quasars\footnote{Hereafter, we will use the word ``quasars'' to refer to AGN brighter than $L_\mathrm{bol}\sim 10^{44}-10^{45}\,\ergs$.} in the Universe. Most large-scale hydrodynamical simulations extend to lengths of $\sim$50-300 cMpc on a side: such volumes are large enough to reproduce statistical samples of massive galaxies as well as faint AGN/quasars, but do not encompass the population of bright quasars, which are simply too rare to be produced in these simulated volumes. For this reason, most comparisons between the quasar (SMBH) populations in simulations and observations mainly focused on the faint (low-mass) end of the distributions \citep{Habouzit2021}.

While simulations struggle to reach volumes large enough to produce statistical samples of bright quasars, on the observational side, wide-field spectroscopic surveys such as the Sloan Digital Sky Survey \citep[SDSS,][]{york2000}, the 2dF QSO redshift survey \citep[2QZ,][]{croom2004}, the Baryon Oscillation Spectroscopic Survey \citep[BOSS,][]{dawson2013}, and the extended Baryon Oscillation Spectroscopic Survey \citep[eBOSS,][]{dawson2016},  have been delivering a comprehensive picture of UV-bright quasars across cosmic time. A basic metric for characterizing this population is the quasar luminosity function (QLF). The QLF describes the comoving number density of quasars as a function of luminosity and has been explored, e.g., across multiple wavelengths and redshift ranges \citep{Boyle2000, Richards2006, Ross2013, Akiyama2018}. The QLF typically follows a broken power-law form, with the break luminosity and the slope of the bright end showing significant evolution with redshift. This evolution reflects the cosmic history of SMBH growth, peaking at $z\sim 2-3$ and declining toward both higher and lower redshifts \citep{Kulkarni2019}. 

An important caveat concerning the QLF is that obscuration of quasars by surrounding dust and gas introduces additional complexity in modeling the observable QLF. Observations suggest that the fraction of obscured AGN varies with both luminosity and redshift \citep[e.g.,][]{Ueda2014, Buchner2015}, potentially affecting comparisons between simulated intrinsic bolometric luminosities and observational measurements in specific bands. Recent efforts to combine X-ray, optical, infrared, and radio observations have begun to provide a more complete census of the AGN population \citep[e.g.,][]{shen2020}, offering more robust benchmarks for simulation predictions.

In addition to the QLF, another fundamental metric that captures key properties of the quasar population in a cosmological context is the clustering of quasars, which serves as a useful probe of the connection between quasars and their host dark matter haloes. Large spectroscopic surveys have measured quasar clustering across various redshift ranges, revealing that quasars typically reside in haloes with masses of $\sim 10^{12-13}\,\msun$, largely independent of luminosity and with modest evolution with redshift, at least up to $z\approx2-3$ \citep[e.g.,][]{porciani_2004,croom2004,Porciani2006,Shen_2006,da_angela2008,Ross2009,White2012,Eftekharzadeh_2015}. These measurements constrain the relationship between quasar activity and halo properties, the typical lifetime of quasars, and the scatter in the SMBH-halo mass relation \citep[e.g.,][]{martini2001, haiman_hui2001, pizzati2024a}. 

Reproducing quasar clustering in simulations is especially challenging, for two key reasons: most clustering measurements focus solely on luminous quasars, as it is much easier to obtain complete and pure samples when considering only bright ($L_\mathrm{bol}\gtrsim 10^{45}-10^{46}\,\ergs$) objects; on top of that, reliable auto-correlation measurements can only be obtained when a sample of $\gtrsim 10^2-10^3$ objects is available. Given the limited box sizes of hydrodynamical simulations, reaching the same comoving volumes probed by surveys of UV-bright quasars is often unfeasible, and a proper comparison between the clustering of quasars in observations and simulations is yet to be achieved. Previous theoretical efforts exist, but have been focused on specific regimes such as low-luminosity sources \citep[][]{Degraf2012, Degraf2017} and very small scales \citep{bhowmick2019}, or have employed dark-matter-only simulations in conjunction with semi-analytic or empirical models \citep[e.g.,][]{fanidakis2013,oogi2016,knox2025}.

In the last few years, new state-of-the-art cosmological simulations have been developed, extending the regimes probed by hydrodynamical simulations to Gpc-scale box lengths, albeit at the cost of using lower numerical resolutions compared to smaller-volume simulations. These include FLAMINGO \citep{FLAMINGO}, MillenniumTNG \citep{pakmor2023}, and Magneticum \citep{dolag2025}.
Thanks to their larger volumes, these simulations offer a new way to probe rare objects such as bright quasars, exploring their distribution up to very large luminosities $L_\mathrm{bol}\ \gtrsim10^{47}\ergs$, and compiling large statistical samples which can be used for clustering studies. 

In this paper, we use the FLAMINGO cosmological simulation \citep{FLAMINGO, Kugel}, which provides an extremely large box size of $L=2.8\,\cGpc$, to perform an in-depth analysis of both the luminosity function and clustering of quasars. We aim to determine whether FLAMINGO can match the observed properties of bright quasars across cosmic history, and we discuss the insights into black hole accretion, growth, and feedback that can be drawn from our comparison with the data.

This paper is structured as follows: Section \ref{sec:method} outlines the FLAMINGO simulation framework as well as the relevant numerical methods and physical prescriptions. In Section \ref{sec:results}, we present our results for the luminosity function and the clustering of quasars. We discuss these results and analyse their consequences in Section \ref{sec:discussion}. Section \ref{sec:conclusions} summarizes our results and discusses potential avenues for future research.

\section{The FLAMINGO simulations} \label{sec:method}

\textsc{FLAMINGO} \citep[Full-hydro Large-scale structure simulations with All-sky Mapping for the Interpretation of Next Generation Observations,][]{FLAMINGO} is a suite of cosmological hydrodynamical simulations developed for cosmology and galaxy cluster physics.
It comprises three different resolutions (Table \ref{tab:runs}), which are all calibrated to the same data, using a machine learning method based on Gaussian process emulators \citep{Kugel}. The two flagship runs use volumes of $(2.8\,\mathrm{cGpc})^3$ and $(1.0\,\mathrm{cGpc})^3$ and baryonic particle masses of $1\times 10^9\,{\rm M}_{\odot}$ and $1\times 10^8\,{\rm M}_{\odot}$ respectively. With $(5040)^3$ baryonic particles, $(5040)^3$ cold dark matter particles and $(2800)^3$ neutrino particles, the $(2.8\,\mathrm{cGpc})^3$ simulation is the largest ever hydrodynamical simulation run to $z=0$. Relevant to our analysis, it also represents one of the biggest cosmological volumes that have ever been simulated with the inclusion of baryonic physics.

The fiducial cosmological parameters in \textsc{FLAMINGO} are taken from the Dark Energy Survey year three \citep[DES Y3;][]{Abbott_2022}. 
Besides the fiducial model, \textsc{FLAMINGO} includes eight astrophysics variations and four cosmology variations, all in $(1.0\,\mathrm{cGpc})^3$ volumes with intermediate resolution (Cosmological parameters for different simulations can be found in Tables \ref{tab:numerical_parameters}, and subgrid model parameters for each astrophysical run can be found in \citet{FLAMINGO}).

In our work, we will primarily employ the $\mathrm{L2p8\_m9}$ run (Table \ref{tab:runs}), as it provides the largest volume at a medium resolution, and such a large volume is needed to study the population of rare bright quasars. In the Appendix, we also explore our results in the context of different runs: we examine the convergence of our results by varying box size and resolution (Appendix \ref{appendix:convergence}), as well as the effects of different cosmological and astrophysical parameters (Appendix \ref{appendix:variations}). Cosmological and astrophysical variations are only available for the fiducial $\mathrm{L1\_m9}$ run, so we use that run (which has a smaller box of $L=1\,\cGpc$) as our benchmark in the Appendixes. 

\begin{table}
    \centering
    \caption{Parameters for differents runs in \textsc{FLAMINGO}: $\mathrm{N_{baryon}}$, $\mathrm{N_{dm}}$, and $\mathrm{N_{\nu}}$ represent the number of baryon particles, dark matter particles, and neutrino particles, respectively.}
    \label{tab:runs}
    \begin{tabular}{llrrr}
    \hline
    Name & Boxsize$\mathrm{(Gpc)}$ & $\mathrm{N_{baryon}}$ & $\mathrm{N_{cdm}}$ & $\mathrm{N_{\nu}}$ \\ \hline
    L1\_m10 &\quad\quad 1 & $900^3$ & $900^3$  & $500^3$\\
    L1\_m9 &\quad\quad 1 & $1800^3$ & $1800^3$  & $1000^3$\\
    L1\_m8 &\quad\quad 1 & $3600^3$ & $3600^3$  & $2000^3$\\
    L2p8\_m9 &\quad\quad 2.8 & $5040^3$  & $5040^3$  & $2800^3$\\
    \hline
    \end{tabular}
\end{table}

\subsection{Simulation setup}

The FLAMINGO simulations were performed using the open source code \textsc{Swift} \citep{swift_2023}. The hydrodynamical equations are solved using the \textsc{sphenix} flavour of smooth particle hydrodynamics (SPH) \citep{SPH_sphenix_2022}. A single massive neutrino species with a mass of $0.06\ \mathrm{eV}$ and two massless species -- that is equal to the minimum allowed by ground-based neutrino oscillation experiments -- are included in \textsc{FLAMINGO}.
Massless neutrinos are accounted for in the calculation of the Hubble expansion rate and in the initial conditions, but are otherwise modeled as a smooth component. The massive neutrinos in \textsc{FLAMINGO} are modelled using $\delta f$ method which is designed to reduce shot noise \citep{elbers_optimal_2021}. The initial conditions are set up using a modified version of \textsc{Monofonic} code, which implements the effects of massive neutrinos (\citealt{hahn_higher_2021}, \citealt{elbers_higher_2022}). 

Haloes and substructures are identified with $\mathrm{VELOCI}_\mathrm{RAPTOR}$ \citep{Elahi_2019}, and their properties are derived using the Spherical Overdensity and Aperture Processor (SOAP, \citealt{McGibbon2025}). For most of our work, we rely on the \texttt{BoundSubhaloProperties} catalog from SOAP. 
These properties are computed for each subhalo identified by $\mathrm{VELOCI}_\mathrm{RAPTOR}$, irrespective of whether it is a field halo, a satellite, or a higher-order substructure. They are based only on the particles that are gravitationally bound to the subhalo, and therefore provide a consistent definition for both central and satellite haloes\footnote{When computing the quasar-host mass functions (QHMFs) in Sec. \ref{sec:discussion_clustering}, we only include central haloes and therefore adopt the more common $M_{\rm 200c}$ halo mass definition, based on the spherical overdensity (SO) catalog from SOAP.}.

As is the case for all the large-volume cosmological simulations, key physical processes that are relevant to black hole seeding, growth, and the evolution of galaxies, are largely unresolved in FLAMINGO and accounted for by subgrid models. 
As mentioned before, the calibration of the subgrid physics is done by machine learning and targets two critical observables: the present-day galaxy stellar mass function and the gas mass fraction in galaxy clusters. As discussed in \citet{Kugel}, the calibration process for subgrid parameters uses a systematic, Bayesian approach (\citealt{Bower2017}, \citealt{Rodrigues2017}) instead of traditional trial-and-error methods. Gaussian process emulators \citep{Rasmussen2004} trained on 32-node Latin hypercubes of simulation are used to give the prediction of different observables. The calibrated subgrid model parameters for each astrophysical run can be found in \citet{FLAMINGO}. 

The subgrid models most relevant to our work include star formation, stellar and AGN feedback, and black hole seeding, accretion, and dynamics.
In FLAMINGO, the multiphase ISM is modeled following \citet{Schaye2008}, and gas particles are converted into star particles using the pressure-dependent star formation rate from \citet{Schaye2008}. Stellar energy feedback is implemented by kinetically kicking the SPH neighbors of young star particles, following the approach of \citet{Chaikin2023}. AGN feedback and the treatment of SMBH evolution are described in more detail below.

\subsection{BH seeding and accretion}

Following \citet{Springel2005}, \citet{Di_Matteo_2008} and \citet{Booth2009}, in \textsc{FLAMINGO}, seed black holes are placed in haloes that are sufficiently massive and do not yet contain a black hole. The haloes are found by running a friend-of-friend (FoF) halo finder\footnote{We note that the on-the-fly FoF halo finder used for black hole seeding is different from the main halo finder that is run on the simulation snapshots, which is based on the $\mathrm{VELOCI}_\mathrm{RAPTOR}$ algorithm \citep{Elahi_2019}.} with linking length $0.2$ times the mean inter-particle distance at regular intervals of $\Delta \log_{10}a=1.00751$. The minimum halo mass for seeding is set to $2.757\times 10^{11}{\rm M}_{\odot}$. The seed of the black hole is positioned at the location of the densest gas particle within the halo. This particle is then transformed into a collisionless black hole particle, assuming the mass of the original gas particle. Black hole processes are computed using the subgrid black hole mass, which is set to be $10^5\,{\rm M}_{\odot}$, whereas the gravitational force is computed using the seed particle mass.

In cosmological simulations, dynamical friction is often underestimated due to insufficient resolution, which prevents the accurate capture of essential interactions around black holes. As a remedy, black holes in FLAMINGO are manually repositioned following the methods proposed by \citet{Springel2005} and \citet{Bahe2022}. This approach involves moving the black hole to the location of the SPH neighbor with the lowest gravitational potential within three gravitational softening lengths, effectively enhancing black hole growth through gas accretion and mergers, thereby improving the efficiency of AGN feedback.

BHs grow through two main modes: gas accretion and mergers.
In \textsc{FLAMINGO}, black holes merge if they are separated by less than $3$ gravitational softening lengths, $r<3\epsilon$, and if their relative velocity satisfies $\Delta v<\sqrt{2GM_\mathrm{BH}/r}$, where $M_\mathrm{BH}$ 
is the mass of the most massive of the two black holes and $r$ is the separation. When a merger occurs, the momentum, subgrid mass, and particle mass of the lower-mass SMBH are transferred to the more massive one, and then the former is removed from the simulation. This approach ensures a clear tracking of SMBH main progenitors over time. In rare cases where multiple SMBHs are eligible to merge within the same time step, each SMBH is swallowed only once by the most massive SMBH for which it is eligible. If an SMBH is set to merge with another that is itself merging simultaneously, it remains in the simulation until at least the next time step.

Following \citet{Springel2005}, in \text{FLAMINGO} black holes accrete at a modified Bondi-Hoyle rate
\begin{equation}
\dot{M}_\mathrm{B}=\alpha \frac{4\pi G^2 M_\mathrm{BH}^2 \rho}{(c_s^2+v^2)^{3/2}},
\label{eq:Bondi-Hoyle}
\end{equation}
where $\rho$ and $c_s$ are the gas density and the speed of sound of the ambient medium, $v$ is the velocity of the black hole with respect to its environment, and the coefficient $\alpha$ is a boost factor which is introduced because the simulations do not resolve the Bondi radius and do not model the multiphase ISM \citep{Booth2009}:
\begin{equation}
\alpha= \begin{cases}1 & n_{\mathrm{H}}<n_{\mathrm{H}}^* \\ \left(\frac{n_{\mathrm{H}}}{n_{\mathrm{H}}^*}\right)^{\beta_{\mathrm{BH}}} & n_{\mathrm{H}} \geq n_{\mathrm{H}}^*\end{cases},
\end{equation}
where $n_{\mathrm{H}}$ represents the hydrogen number density, and $n_{\mathrm{H}}^*$ is a critical density threshold that distinguishes different gas phases. The parameter $\beta_{\mathrm{BH}}$ controls how strongly the boost factor increases with gas density when $n_{\mathrm{H}} \geq n_{\mathrm{H}}^*$, accounting for the unresolved cold interstellar medium and ensuring efficient black hole growth in high-density environments.

The accretion rate is also assumed to be limited by the Eddington rate:
\begin{equation}
\dot{M}_{\mathrm{Edd}}\equiv \frac{4\pi GM_{\mathrm{BH}}m_{\mathrm{p}}}{\epsilon_{\rm r} \sigma_{\mathrm{T}} c},
\label{eq:edd_rate}
\end{equation}
where $m_{\rm p}$ is the proton mass, $\sigma_{\mathrm{T}}$ is the Thomson cross-section and $\epsilon_{\rm r}$ is the radiative efficiency, which is the ratio between bolometric luminosity $L_\mathrm{bol}$ and the accretion energy rate of a black hole:
\begin{equation}
\epsilon_{\rm r} = \frac{L_\mathrm{bol}}{\dot{M}_{\mathrm{BH} }c^2}.
\end{equation}
The radiative efficiency is assumed to be \(0.1\) in \textsc{FLAMINGO}, broadly in agreement with the average value for radiatively efficient accretion onto a Schwarzschild black hole \citep{Shakura1973}. Given this Eddington limit, the accretion rate reads:
\begin{equation}
\dot{M}_\mathrm{accr} = \mathrm{min}(\dot{M}_{\mathrm{Edd}},\dot{M}_\mathrm{B}).
\label{eq:accretion_rate}
\end{equation}
Gas accretion increases the black hole subgrid mass as 
\begin{equation}
\dot{M}_\mathrm{BH}=(1-\epsilon_{\rm r})\dot{M}_\mathrm{accr},
\label{eq:1}
\end{equation}
and decreases the black hole particle mass by $\epsilon_{\rm r} \dot{M}_\mathrm{accr}$. The decrease of the black hole particle mass accounts for the loss of rest mass to radiation; this implies that the bolometric luminosity of quasars (which we will use in the subsequent sections) can be simply expressed as: 
\begin{equation}
L_\mathrm{bol}=\epsilon_{\rm r} \dot{M}_\mathrm{accr} c^2=\frac{\epsilon_{\rm r} \dot{M}_\mathrm{BH}}{1-\epsilon_{\rm r}}c^2.
\label{eq:luminosity}
\end{equation}

\subsection{AGN feedback}

For the fiducial model, AGN feedback is applied thermally, as described by \citet{Booth2009}. However, in some of the intermediate-resolution runs, anisotropic, kinetic feedback \citep{husko2023} is employed to assess the sensitivity of the results to different implementations of AGN feedback. 

In the thermal implementation, a fraction of the accreted rest-mass energy available in each time step, 
\(\epsilon_{\mathrm{f}}\epsilon_{\mathrm{r}} \dot{M}_{\mathrm{accr}} c^2 \Delta t\), 
is coupled to the surrounding gas. Because the energy must be sufficient to heat at least the mass of one gas particle, the resulting post-shock gas temperature is therefore underestimated, which in turn leads to an overestimation of radiative cooling rates \citep{Dalla2012}. In order to avoid the overcooling problem, energy is stored in a subgrid reservoir of black hole particle until it suffices to heat a fixed number (\(n_{\mathrm{heat}}\)) of neighbouring gas particles by a predefined temperature increment (\(\Delta T_{\mathrm{AGN}}\)), instead of injecting the feedback energy at every time step. The fiducial model adopts \(n_{\mathrm{heat}} = 1\). The feedback efficiency is set to \(\epsilon_{\mathrm{f}} = 0.15\) \citep{Booth2009}, such that \(\epsilon_{\mathrm{f}}\epsilon_{\mathrm{r}} = 0.015\) of the accreted rest-mass energy is used for feedback. \(\Delta T_{\mathrm{AGN}}\) is treated as a calibrated parameter that controls the burstiness and efficiency of AGN heating.

In the kinetic feedback runs, AGN feedback is instead implemented in a jet-like fashion \citep{Hu_ko_2022}. This model is motivated by the expectation that, at low Eddington ratios, accretion flows become radiatively inefficient and geometrically thick, which favours the launching of collimated relativistic jets \citep[e.g.,][]{Yuan2014}. The \textsc{FLAMINGO} jet simulation is a simplified version of the black hole spin
and AGN jet implementation of \citet{Hu_ko_2022}. \textsc{FLAMINGO} employs the jet mode at all accretion rates and adopts a constant feedback efficiency of $0.015$, consistent with the fiducial thermal model. In \textsc{FLAMINGO}, the jet orientation is determined by the black-hole spin vector, which evolves in response to the angular momentum of the gas within the black-hole smoothing kernel, as well as BH mergers, jet spindown, and Lense–Thirring torques. Although it is a strong assumption given the limited resolution of the simulation, jet redirection is expected to occur only on Gyr time-scales, much longer than the duration of individual jet episodes, so its effects on long-term feedback are negligible \citep{Hu_ko_2022}. Jets are launched once the black hole energy reservoir exceeds $2 \times \tfrac{1}{2} m_{\mathrm{g}} v_{\mathrm{jet}}^2$ by kicking two gas particles from within the black hole’s SPH kernel. The two particles closest to the spin axis are then accelerated along directions chosen randomly within cones of 7.5$^\circ$ opening angle on either side of the axis. Since real AGN jets have quasi-relativistic velocities that cannot be resolved at the available resolution, the jet velocity $v_{\mathrm{jet}}$ is treated as a subgrid parameter, analogous to $\Delta T_{\mathrm{AGN}}$ in the thermal model.

\section{Results} \label{sec:results}

In this section, we present the predictions for the QLF and the clustering of quasars as derived from \textsc{FLAMINGO}, and compare these findings with observational data.

\subsection{Quasar luminosity function} \label{sec:qlf}

We focus on a luminosity range from $10^{42}$ to $10^{48} \,\mathrm{erg\ s^{-1}}$, which encompasses the range of SMBHs resolved by FLAMINGO (Appendix \ref{appendix:convergence}) and whose number density is large enough to be represented in the simulated box. We split this range into several bins in logarithmic space. We define the interval of each bin as $\Delta \log_{10}(L_\mathrm{bol})$. We then compute the bolometric luminosity of quasars following \cref{eq:luminosity}, and count the number of quasars in each bin ($N_i$) to compute the QLF, which can be expressed as: 
\begin{equation}
\phi(L_{\mathrm{bol},i}) = \frac{N_i}{\mathrm{(box~size)}^{3}\times \mathrm{\Delta \log_{10}}(L_\mathrm{bol})}.
\end{equation}
Uncertainties on the QLF are calculated using Poisson statistics.

\begin{figure*}
    \centering
    \includegraphics[width=1.0\textwidth]{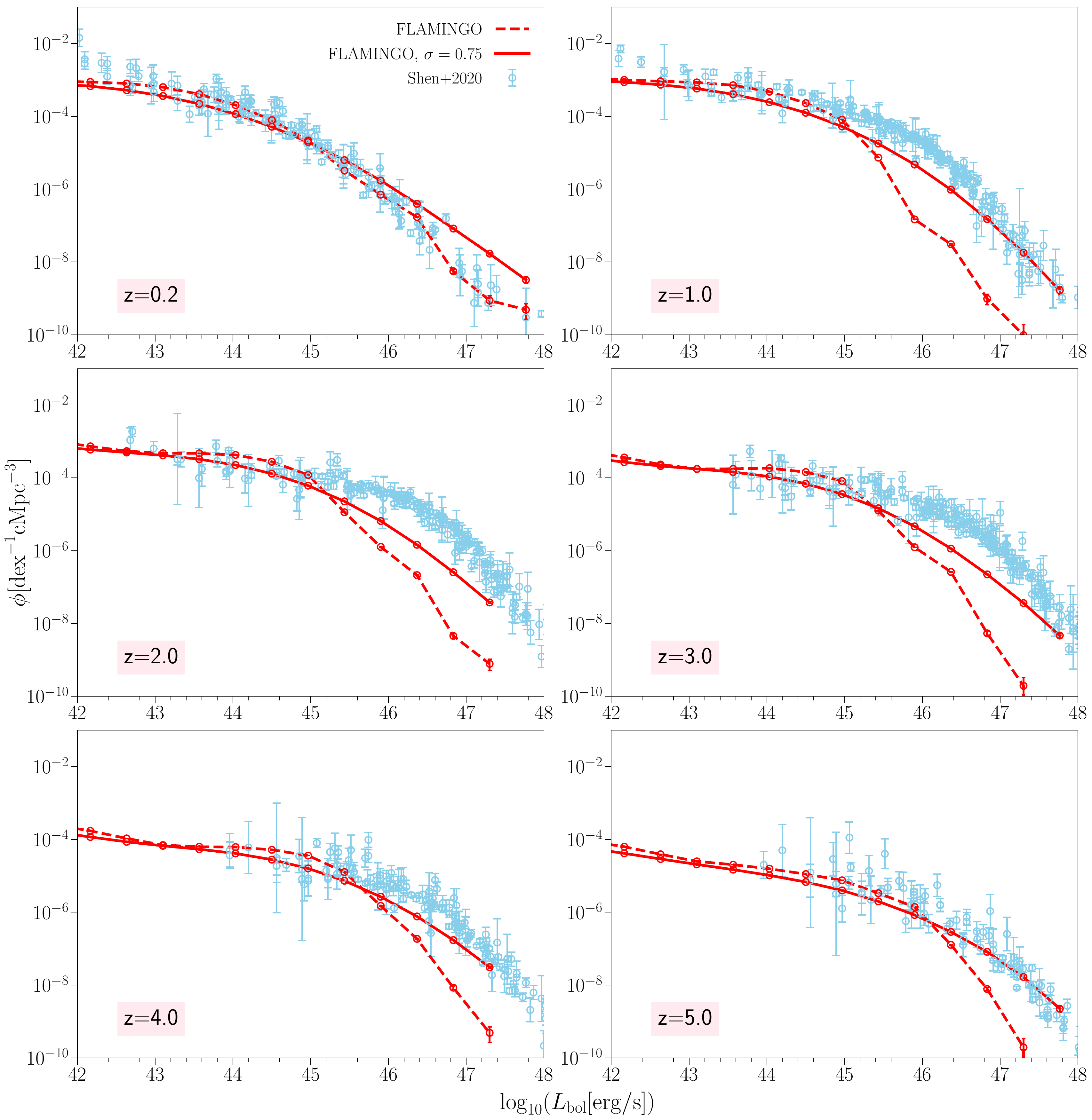} 
    
    \caption{Comparison of the bolometric quasar luminosity function (QLF) computed for the FLAMINGO L2p8\_m9 run (red lines) with observational data compiled by \citet{shen2020} (blue points) for different redshifts. The red dashed lines represent the QLF obtained directly from the FLAMINGO catalogues (see \cref{sec:method}), while the red solid lines represent the effect of adding log-normal scatter (with $\sigma=0.75$ dex) to the bolometric luminosity of quasars.
    \label{fig:QLF}}
\end{figure*}

\cref{fig:QLF} shows the QLF at six different redshifts from $z=0.2$ to $z=5.0$, alongside observational data. The red dashed lines represent the QLF computed using the L2p8\_m9 run of \textsc{FLAMINGO}, while the blue points are taken from the \citet{shen2020} bolometric luminosity function compilation. These data include optical/UV, X-ray, and infrared (IR) detected AGN. The optical/UV data ($2500\,\mathrm{\AA} \leq \lambda \leq 1\,\mathrm{\mu m}$ for optical and $600\,\text{\AA} \leq \lambda \leq 2500\,\text{\AA}$ for UV) were primarily sourced from the compilations of \citet{Hopkins2007}; \citet{Giallongo2012}; \citet{Manti2016}; and \citet{Kulkarni2019}. In the X-ray, in addition to the observations compiled in \citet{Hopkins2007}, observational data from \citet{Ebrero2009}; \citet{Aird2008}; \citet{Silverman2008}; \citet{Yencho2009}; \citet{Aird2010}; \citet{Fiore2012}; \citet{Ueda2014}; \citet{Aird2015a, Aird2015b}; \citet{Miyaji2015}; \citet{Khorunzhev2018} are included. “IR” wavelengths are defined as $\lambda \geq 1 \mathrm{\mu m}$. In the IR, in addition to the observations compiled in \citet{Hopkins2007}, observations from \citet{Assef2011} and \citet{Lacy2015} are included. More details about the data compilation can be found in \citet{shen2020}.

We find that, at low redshifts ($z=0.2$, top left panel), the simulation results are remarkably consistent with observations. We note that this agreement is highly non-trivial, since FLAMINGO was not explicitly calibrated on quasar luminosities. Most simulations reproduce the local black hole mass-stellar mass relation by implicitly tuning AGN feedback parameters \citep[e.g.,][]{Habouzit2021}, but the fact that FLAMINGO also recovers the low-$z$ QLF indicates that the corresponding SMBH accretion rates arise naturally from the underlying galaxy formation model. Toward cosmic noon, however, the agreement weakens: FLAMINGO reproduces the quasar number densities only at the faint end of the QLF ($L_\mathrm{bol}\approx 10^{43}-10^{45}\,\ergs$). At higher luminosities ($L_\mathrm{bol}\gtrsim 10^{45}\,\ergs$), the simulated number density declines much more steeply than observed, leading to a strong underestimation of the brightest quasars ($L_\mathrm{bol}\gtrsim 10^{46}-10^{47}\,\ergs$). At higher redshifts ($z\gtrsim4$), this discrepancy is less pronounced, and the simulation predictions only begin to diverge noticeably from the data beyond $L_\mathrm{bol}\gtrsim 10^{46}\,\ergs$.

Given its relatively low resolution (baryonic particle mass of $1\times10^9\,\msun$ for the L2p8\_m9 run) and the fact that black hole accretion is capped at the Eddington limit, it is perhaps not surprising that FLAMINGO struggles to reproduce the observed number density of bright quasars. The processes driving the accretion of gas on SMBHs and the evolution of quasar luminosities over time, in fact, are all sensitive to the small-scale physics that drives the gas from the central regions of the galaxy down to the sphere of influence of the SMBH. While a low-resolution simulation can hope to describe effectively the galactic-scale processes that ultimately determine the evolution of the SMBH over cosmic timescales, it has no hope in resolving the small-scale variability of the SMBH accretion process that is the primary driver of the quasar's instantaneous luminosity. Hence, it is very likely that the scatter in the luminosity of quasars in the simulation is significantly underestimated. For this reason, we experiment by adding log-normal scatter to the bolometric luminosities of quasars in FLAMINGO, and study the effect of this addition on the resulting quasar properties\footnote{The addition of log-normal scatter does not conserve the mean of the QLF distribution. To correct for this, we conserve the total luminosity emitted by all quasars by manually setting the mean of the sampled log-normal distribution to zero.}. When adding scatter, the bolometric luminosities of quasars can exceed the Eddington limit. Therefore, the inclusion of scatter leads to apparent super-Eddington luminosities, without modifying the Eddington-limited SMBH growth implemented in \textsc{FLAMINGO} (\crefrange{eq:Bondi-Hoyle}{eq:accretion_rate}). Physically, this scatter means that the actual, instantaneous SMBH growth may fluctuate relative to the growth modeled by the simulation.

We note that our aim is not to determine the precise level of scatter that best matches each of the datasets considered, nor to suggest that scatter is the primary cause of the discrepancy between model and observations. Instead, we seek to assess, through a simple correction, the impact of enhanced variability in quasars’ instantaneous bolometric luminosities, as expected given the limited resolution of FLAMINGO. To this end, we adopt a representative log-normal scatter of $\sigma = 0.75$ dex throughout this work, where $\sigma$ denotes the standard deviation of $\log_{10} L_{\rm bol}$. This relatively large value should be regarded as a conservative upper limit, ensuring that the two cases considered -- “with” and “without” scatter -- bracket the plausible range of outcomes. Larger scatter values would likely overproduce bright quasars and substantially distort the shape of the QLF.

The red solid lines in \cref{fig:QLF} show the QLFs at different redshifts when a 0.75 dex log-normal scatter in quasar luminosity is included. As expected, the faint-end number density decreases slightly compared to the no-scatter case, but remains broadly consistent with observational constraints. At the bright end, by contrast, the steep decline of the no-scatter QLF is considerably flattened, bringing the predictions into closer agreement with the data—except at $z=0.2$, where the no-scatter model already provides an excellent fit and the inclusion of scatter leads to a modest overprediction of bright quasars. Nonetheless, a clear discrepancy persists between the simulated and observed QLFs, particularly around cosmic noon ($z \approx 1\text{–}3$). We therefore conclude that, even after accounting for the likely underestimated intrinsic scatter in quasar luminosities, the simulation can only partially capture the global evolution of quasar activity across cosmic history.

\begin{figure*}

	\includegraphics[width=1.0\textwidth]{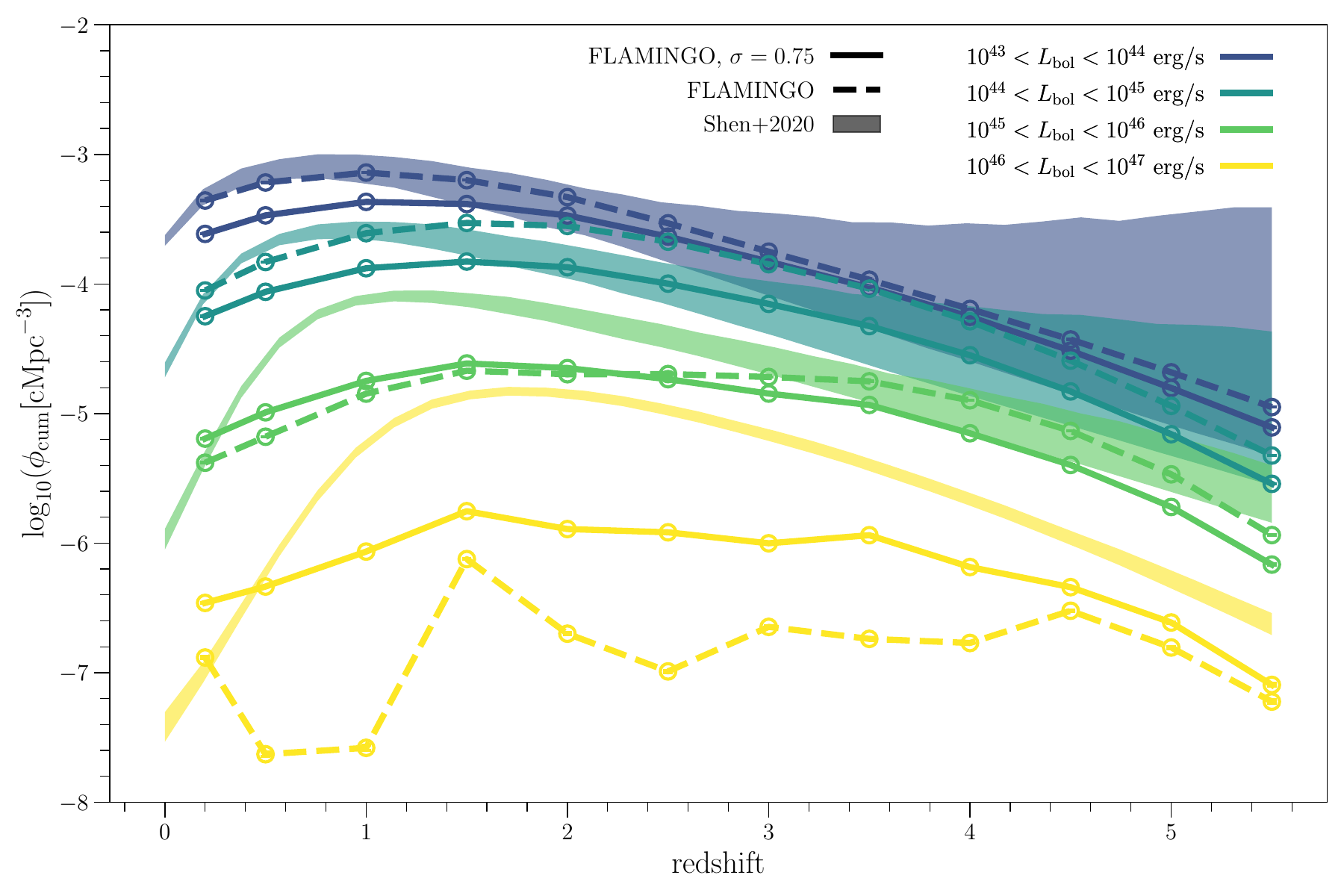}
    \caption{Cumulative Quasar Luminosity Function (cQLF) in different bolometric luminosity bins. The colored lines represent the predictions from the \textsc{FLAMINGO} L2p8\_m9 simulation with (solid) and without (dashed) 0.75 dex of additional scatter in bolometric luminosity. The shaded region represents the range between the maximum and minimum values obtained by reproducing the constraints from ``Global Fit A'' and ``Global Fit B'' taken from the \citet{shen2020} model of observational data at different redshifts.}
    \label{fig:cumu}
\end{figure*}

This can be effectively visualized by looking at the cumulative quasar luminosity function (cQLF) as a function of redshift. The cQLF can be obtained by integrating the QLF in a given luminosity range. In practice, for the simulation, we can directly count the number of quasars in each luminosity bin at different snapshots (corresponding to different redshifts). For the observational data, instead, we employ the parametrization of the global evolution of the QLF proposed by \citet{shen2020}. They introduced two different global fitting approaches: the first (“Global Fit A”) allows the faint-end slope to evolve flexibly with redshift, capturing potential complexities in its evolution, while the second (“Global Fit B”) imposes a smoother, monotonic evolution that enforces a more regular trend. The detailed functional forms and fitting methodology are reported in Appendix~\ref{app:shenfits}. In \cref{fig:cumu}, we highlight with shaded colors the regions between the minimum and the maximum constraints obtained by putting together the two models. Each color represents a different bin in bolometric luminosity. 

We compare the constraints coming from observational trends with the predictions obtained from FLAMINGO. The solid and dashed lines in \cref{fig:cumu} represent the cQLFs for the simulation with and without scatter, respectively. 
We find that the faint end of cQLF is well reproduced by the simulation, with a consistent overall trend and a peak around cosmic noon. Adding scatter brings the results in closer agreement to observations for $z\gtrsim2$, but reduces the number of quasars as compared to the data predictions at lower redshifts. However, the impact of scatter is generally minor for faint ($L_\mathrm{bol}\lesssim10^{45}\,\ergs$) quasars. 

In the bright luminosity bins, on the other hand, the difference between FLAMINGO and the data becomes larger, with the simulation underestimating the number density of very bright quasars by up to 3-4 orders of magnitude. This is because the overall evolution of bright quasars in FLAMINGO does not follow the established observational trend: the peak at cosmic noon is only marginally present for $10^{45}\,\ergs<L_\mathrm{bol}<10^{46}\,\ergs$, and disappears completely for the brightest bin. Adding scatter significantly reduces the gap between the simulation and observations, enhancing the predicted number density at cosmic noon, but it is not sufficient to match the steep rise and fall of the bright quasar population as seen in observations across cosmic epochs. 

Finally, we have verified that our results are robust against both numerical choices and model variations. Convergence tests with different box sizes and resolutions (Appendix~\ref{appendix:convergence}) confirm that the QLFs are well converged across the luminosity range probed by current observations, with noticeable differences only at the faintest luminosities where observational constraints are limited. Likewise, variations in cosmology or subgrid astrophysics have only minor effects on the QLF. Modest differences arise, for instance, in the low-$\sigma_8$ cosmology, which predicts a reduced quasar number density at high $z$, and in the kinetic AGN feedback runs, which tend to boost the abundance of bright sources. The results for the kinetic (jet) feedback runs are especially interesting, as they bring the AGN abundance predicted by FLAMINGO closer to the observed data (Fig. \ref{fig:phy}). A plausible reason for this is that kinetic feedback generally operates on longer timescales, and it is therefore less immediately effective at quenching black hole accretion than thermal feedback, leading to an enhanced number of luminous AGN. Further discussion on this and on the results for other astrophysical and cosmological variations is provided in Appendix~\ref{appendix:variations}.

\subsection{Quasar clustering}

In addition to studying the relative number of quasars of different luminosity, we are also concerned with their spatial distribution, so we investigate their clustering. The most basic statistic that can describe the clustering of a population of objects is the two-point auto-correlation function.
In cosmological simulations with periodic boxes, the correlation function can be obtained using a simple pair-counting estimator: 
\begin{equation}
\xi(r)=\frac{\mathrm{DD}(r)-\mathrm{RR}(r)}{\mathrm{RR}(r)},
\end{equation}
where $\mathrm{RR}(r)$ and $\mathrm{DD}(r)$ represent the number of random-random (i.e., randomly distributed) and quasar-quasar pairs computed for different bins of radius $r$, respectively.
We estimate the number of pairs, $\mathrm{DD}(r)$,  using the \textsc{CorrFunc} package \citep{Sinha2019,Sinha2020}. For periodic boxes, the number of random pairs can be computed analytically for a choice of radial bins using the following equation:

\begin{equation}
\mathrm{RR}_i = \frac{4\pi}{3}(r_{i+1}^3-r_{i}^3)\frac{N^2}{L^3},
\label{eq:RR}
\end{equation}
where $\{r_i\}$ represent the bin edges, $N$ is the total number of quasars in the box, and $L$ is the size of the box.

We compute the uncertainty in the two-point correlation function by simply assuming Poisson uncertainties on the pair counts: 
\begin{equation}
\Delta \xi(r) = \frac{\sqrt{\mathrm{DD}(r)}}{\mathrm{RR}(r)}.
\label{eq:ep}
\end{equation}

The three-dimensional correlation function $\xi(r)$ is not a direct observable, because the line-of-sight dimension can only be observed in redshift space, and the conversion from redshift to comoving distance suffers from the effect of redshift-space distortions. For this reason, observations use the projected correlation function $w_{\rm p}(r_{\rm p})$, which quantifies the clustering of objects by integrating the two-point correlation function along the line of sight to average out the effects of redshift-space distortions.

\begin{figure*}
    \centering
    \includegraphics[width=1.0\textwidth]{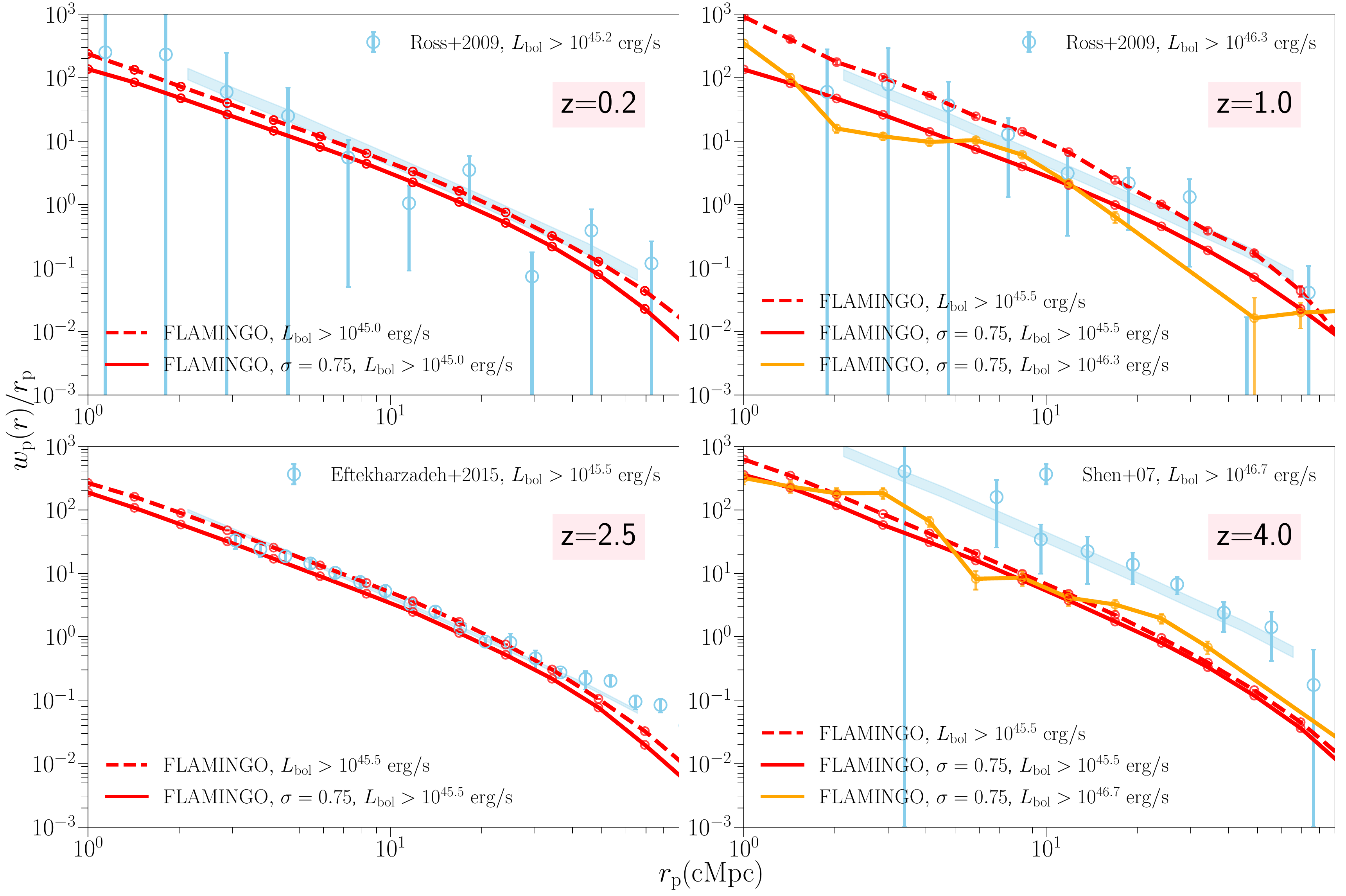} 
    
    \caption{The projected auto-correlation functions of quasars at different redshifts. Solid lines represent the results for the FLAMINGO L2p8\_m9 run with the inclusion of 0.75 dex of scatter in quasar luminosity, while dashed lines show the clustering when scatter is not included. When computing the auto-correlation functions, we only include quasars that are brighter than a luminosity threshold $L_\mathrm{bol,thr}$, that -- whenever possible -- is set close to that used by observations (see main text for details). For \(z=1.0\) and \(z=4.0\), we employ a lower luminosity threshold, but show the results obtained using the same luminosity $L_\mathrm{bol,thr}$ as in the observations -- which becomes possible only in the case with scatter -- with solid orange lines. Blue points show the measured auto-correlation functions of quasars at the same redshifts from SDSS \citep{Ross2009, Shen_2006} and BOSS \citep{Eftekharzadeh_2015}. The shaded blue region highlights the 1-sigma confidence interval for the power-law fit performed on the observational data, and serves as a useful reference in our data-model comparison. 
    \label{fig:clustering}}
\end{figure*}

We follow standard practice and compute the correlation function on a two-dimensional grid of pair separations parallel ($\pi$) and perpendicular ($r$) to the line of sight. Then, the projected correlation function $w_{\rm p}(r_{\rm p})$ can be obtained by integrating $\xi(r_{\rm p}, \pi)$ along the $\pi$ direction, where $\xi(r_{\rm p}, \pi)$ is the correlation function expressed as a function of the transverse separation \(r_{\rm p}\) and the line–of–sight separation \(\pi\):
\begin{equation}
w_{\rm p}(r_{\rm p}) = \int_{-\infty}^{\infty} \xi (r_{\rm p}, \pi)d\pi.
\end{equation}
In practice, we integrate it to a large but finite cutoff value $\pi_\mathrm{max}$. As a result, the projected correlation function can be related to the two-point correlation function via:
\begin{equation}
w_{\rm p}(r_{\rm p}) = 2\int_{r_{\rm p}}^{\pi_\mathrm{max}} \frac{r\xi(r)}{\sqrt{r^2-r_{\rm p}^2}}dr.
\label{eq:wp}
\end{equation}

In \cref{fig:clustering}, we show with red lines the resulting auto-correlation functions of quasars at different redshifts ($z=0.2; 1.0; 2.5; 4.0$), for the cases with (solid) and without (dashed) 0.75 dex of scatter in bolometric luminosity (see \cref{sec:qlf}). 
To obtain the auto-correlation function, we split the range from $1$ to $200\ \mathrm{cMpc}$ into logarithmic bins. Using \textsc{CorrFunc}, we count the number of quasar pairs within each bin and calculate the two-point correlation function using Eq.~\ref{eq:ep}. Then, we obtain the projected correlation function using Eq.~\ref{eq:wp}. We set the value of $\pi_\mathrm{max}$ in a way that is consistent to the observations we compare to at each redshift: $\pi_\mathrm{max}= 100\, \mathrm{cMpc}$ at redshifts $z=0.2, 1.0$, and $4.0$, and $\pi_\mathrm{max}=75\ \mathrm{cMpc}$ at $z=2.5$.

We compare each of the quasar clustering predictions with the autocorrelation functions measured by wide-field spectroscopic surveys such as SDSS and BOSS at different redshifts in \cref{fig:clustering}. The blue points represent the observed auto-correlation functions measured by \citet{Ross2009} at $z=0.08-0.30$ (top left panel) and $z=0.92-1.13$ (top right), \citet{Eftekharzadeh_2015} at $z=2.2-2.8$ (bottom left), and \citet{Shen_2006} $z=3.5-5$ (bottom right). We also highlight with a blue shaded region the 1-sigma uncertainties obtained from the power-law fits to the auto-correlation function data performed by \citet{Ross2009, Eftekharzadeh_2015, Shen_2006}. We use the fits with a constant power-law slope of $-2$ and a varying autocorrelation length parameter, $r_0$.

In order to compute the auto-correlation function in the simulation, we only include quasars brighter than a luminosity threshold, $L_\mathrm{bol, thr}$. Since we aim to reproduce the results of observations, we set this luminosity threshold as close as possible to what was used in the surveys. This implies that we use $L_\mathrm{bol,thr}=10^{45}\,\ergs$ for $z=0.1$ (taken from \citealt{Ross2009}) and $L_\mathrm{bol,thr}=10^{45.5}\,\ergs$ for $z=2.5$ \citep{Eftekharzadeh_2015}. For $z=1$ and $z=4$, we should in principle use higher luminosity thresholds of $L_\mathrm{bol,thr}=10^{46.3}\,\ergs$ \citep{Ross2009} and $L_\mathrm{bol,thr}=10^{46.7}\,\ergs$ \citep{Shen_2006}\footnote{We convert their $\mathrm{M}_{1450}$ magnitude limit to a bolometric luminosity using the conversion from \citet{runnoe2012}; see \citet{pizzati2024a} for more details.}, respectively. However, when scatter is not added, we find that such thresholds yield extremely noisy auto-correlation functions in FLAMINGO, as the simulated volume does not contain enough quasars to measure clustering robustly. For this reason, we set a lower threshold of $L_\mathrm{bol,thr}=10^{45.5}\,\ergs$ for these two redshifts, in agreement with the threshold set at $z=2.5$. Achieving clustering constraints for bright quasars at the precision of wide-field surveys \citep{pizzati2024a} would require much larger simulated volumes ($\gtrsim 100 ,\cGpc^3$). Once scatter is added, however, the abundance of bright quasars increases substantially, enabling us to apply the same luminosity thresholds as in the observations—though the resulting clustering measurements remain noisy and highly uncertain. These results are shown by the orange solid lines in the z=1.0 and z=4.0 panels of \cref{fig:clustering}. We further examine in \cref{sec:discussion_clustering} how quasar clustering depends on luminosity and quantify the impact of varying $L_\mathrm{bol,thr}$ on our results.

The comparison between the measured auto-correlations and our predictions based on FLAMINGO in \cref{fig:clustering} reveals that, in general, the simulation is able to reproduce the clustering of $z\lesssim3$ quasars with a good level of accuracy. At $z\approx0-1$, clustering measurements are quite noisy due to the relative rarity of low-z quasars (see e.g., \cref{fig:cumu}), and the resulting constraints on theoretical models are relatively weak. Both our FLAMINGO predictions (for the cases with and without scatter in quasar luminosity, respectively) broadly fit the data. By using the power-law fits (blue shaded regions) as a reference, we can see that the case without scatter seems to reproduce the data better, especially in the most constraining region between $\approx10-30\,\cMpc$. The $z\approx2.5$ data from the BOSS survey (bottom left panel), on the other hand, are much more constraining, as they focus on a population of quasars at cosmic noon with a broad range of luminosities. Predictions for FLAMINGO reproduce these tight constraints surprisingly well, especially for the case with no added scatter (dashed line). Adding scatter results in a slightly weaker clustering, but the effect is smaller than at $z\lesssim1$, and the resulting auto-correlation function (solid line) only mildly underestimates the observed clustering at all scales. 

At $z\approx4$, the clustering of quasars in FLAMINGO is also very similar in the ``with'' and ``without'' scatter cases for $L_\mathrm{bol,thr}=10^{45.5}\,\ergs$. However, these predictions are in clear disagreement with the clustering measured by \citet{Shen_2006} using SDSS data. The \citet{Shen_2006} data predict a remarkably strong quasar clustering, which implies a rapid evolution of quasar properties with redshift \citep[e.g.,][]{Myers2007}. Many studies have highlighted the challenges that such strong clustering measurement presents for quasar evolution and population models \citep[e.g.,][]{white_2008, shankar_2010, White2012, pizzati2024a}. Additionally, several measurements at similar \citep{he2018} and higher \citep{eilers2024, pizzati2024b} redshifts have challenged the high-$z$ clustering strength and evolution implied by the \citet{Shen_2006} results. It is therefore not surprising that our simulation, together with many other empirical and semi-analytic models \citep[e.g.,][]{conroy_white2013,fanidakis2013,veale2014,aversa2015, trinity}, predicts significantly weaker clustering for quasars at $z\approx4$ than observed by \citet{Shen_2006}. Nonetheless, the study of \citet{Shen_2006} remains founded on the largest and most solid sample of high-$z$ quasars to date, and represents our best knowledge of how quasar clustering evolves beyond cosmic noon. 

In the context of our analysis, we note an important caveat: the comparison between the $z=4$ clustering data and our model is limited by the fact that we can match the observed bolometric luminosity threshold only in the case with scatter. In the absence of scatter, bright quasars in the FLAMINGO simulation are too rare to allow a robust computation of the autocorrelation function. Although the “with” and “without” scatter cases exhibit similar clustering strengths at lower bolometric luminosities (\(L_{\mathrm{bol,thr}} = 10^{45.5}\,\ergs\)), we expect this agreement to break down when we consider brighter sources. This is because the impact of the Eddington bias introduced by scatter increases rapidly toward the bright end of the luminosity function.
Consequently, non-negligible differences between the “with” and “without” scatter cases are expected at high luminosities, a trend that is also reflected in the QHMFs presented in Sec.~\ref{sec:discussion_clustering}. This introduces an uncertainty in our predictions that we cannot address in the context of the current analysis, owing to the intrinsic limitations of the FLAMINGO model. Nevertheless, the markedly different high-$z$ clustering evolution implied by \citet{Shen_2006} compared to various theoretical models remains intriguing. Our results, being in line with these models, further highlight this discrepancy and warrant further investigation.

\begin{figure}
    \centering
    \includegraphics[width=0.49\textwidth]{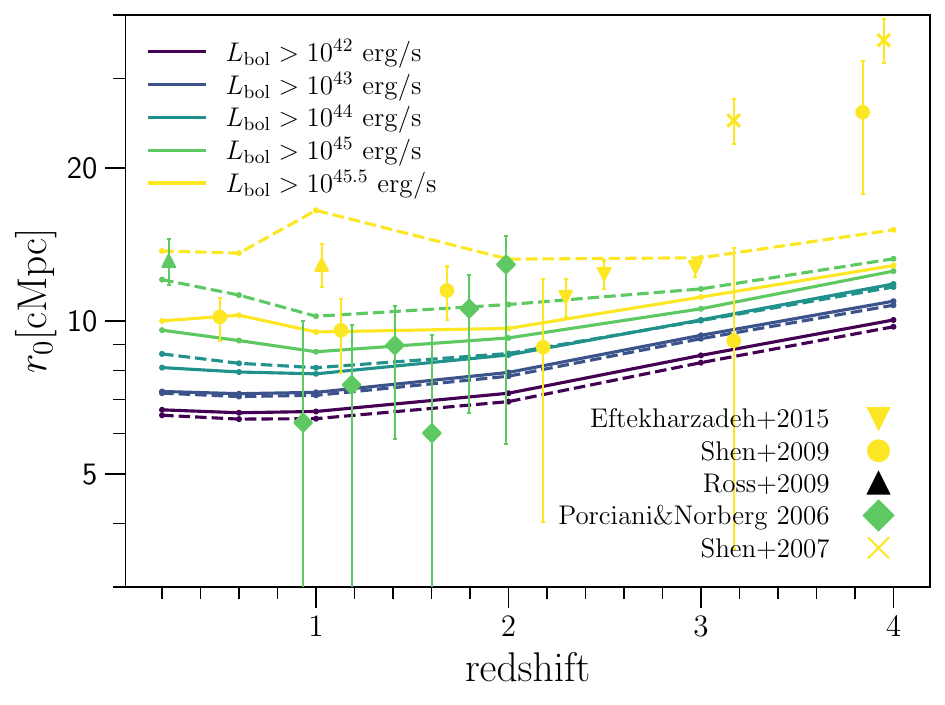}
    
    \caption{Correlation length, $r_0$, as a function of redshift. Colored lines show the predictions from the FLAMINGO L2p8\_m9 simulation for different luminosity thresholds, while data points correspond to observational data for sources with $L_{\rm bol} \gtrsim 10^{45}\, \ergs$ (green) and $L_{\rm bol} \gtrsim 10^{45.5}\, \ergs$ or higher (yellow). The solid (dashed) lines correspond to the case with (without) the addition of 0.75 dex of scatter in quasar luminosity. Correlation length measurements are taken from the $-2$ power-law fits performed by \citet{Eftekharzadeh_2015}, \citet{Shen_2009, Shen_2006}, \citet{Ross2009}, and \citet{Porciani2006}. 
    \label{fig:c_length}}
\end{figure}

To effectively visualize the evolution of quasar clustering with redshift implied by FLAMINGO and compare it to data, we focus on the correlation length, which is defined as the three-dimensional comoving separation \(r_0\) at which the three-dimensional auto-correlation function equals unity, i.e. \(\xi(r_0)=1\). To accurately determine this length, we computed the two-point auto-correlation function of quasars and then derived the correlation length using interpolation methods, ensuring a precise determination of the scale at which the correlation function intercepts unity.
\cref{fig:c_length} shows the predicted correlation length from FLAMINGO as a function of redshift. Different luminosity thresholds, $L_\mathrm{bol,thr}$, are shown with different colors, while the cases with and without scatter in quasar luminosity are shown with solid and dashed lines, respectively.  
We compare these lines to a set of correlation lengths measured from observations \citep{Eftekharzadeh_2015,Shen_2009, Shen_2006,Ross2009,Porciani2006}. We broadly divide the observations according to their luminosity thresholds: green for $L_\mathrm{bol, thr}>10^{45}\ \mathrm{erg\ s^{-1}}$ and yellow for $L_\mathrm{bol, thr}>10^{45.5}\ \mathrm{erg\ s^{-1}}$ and higher. Note that we stop at a luminosity threshold of $10^{45.5}\ \mathrm{erg\ s^{-1}}$ because brighter quasars are rare, leading to a very noisy correlation function and making it impossible to derive an accurate correlation length.  

The FLAMINGO results in \cref{fig:c_length} reveal two clear trends: the clustering of quasars increases with both redshift and luminosity. The increase with redshift appears to get stronger at cosmic noon, as predicted by several theoretical models \citep[e.g.,][]{Hopkins2006}, while at $z\lesssim1$ the redshift dependence is mainly flat (with considerable variations due to the noisy correlation length measurements for the rare low-$z$ quasar population). As for the trend with quasar luminosity, there is a clear difference between the original FLAMINGO runs and those with additional $L_\mathrm{bol}$ scatter. The original runs predict a rather steep, non-linear dependence of clustering with luminosity. The addition of scatter smooths out this dependence, resulting in a milder transition between the fainter and brighter quasar regimes. 
Similar trends with redshift and luminosity are also present in observations, albeit with a large spread due to the different samples and methodologies used by different studies. In general, observations show a mild but steady increase in clustering with redshift from $z\approx0.5$ to $z\approx3$, with a much steeper dependence at higher redshift that is implied by \citet{Shen_2006, Shen_2009, arita2023} but is still controversial (see discussion above). The observed dependence of clustering on luminosity is generally mild or even completely absent \citep[e.g.,][]{adelberger_steidel2005,Porciani2006,Shen_2009,
Eftekharzadeh_2015}, in closer agreement with our predictions from the case with added scatter. For a more detailed discussion on the evolution of clustering with luminosity, we refer the reader to \cref{sec:discussion_clustering}.

As for the QLF, we also check numerical convergence and explore the effects of different cosmological and astrophysical variations on the resulting clustering signal in Appendices \ref{appendix:convergence} and \ref{appendix:variations}.

\section{Discussion} \label{sec:discussion} 

\subsection{Why is the number of bright quasars underestimated in FLAMINGO?} \label{sec:discussion_qlf}

\begin{figure*}
    \centering
    \includegraphics[width=1.0\textwidth]{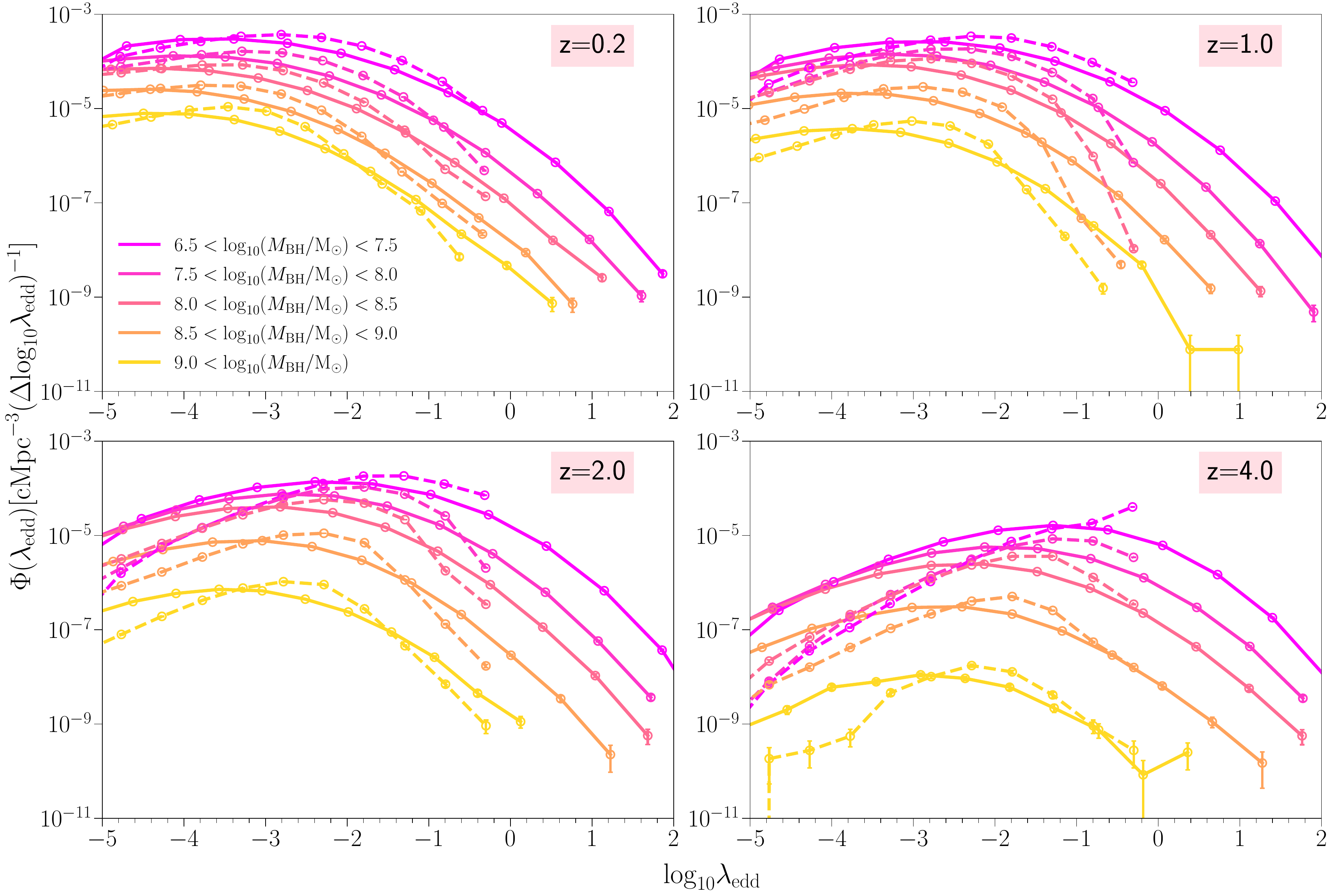} 

    \caption{Eddington ratio distribution functions (ERDFs) split in bins of black hole mass, with different colors representing different black hole mass ranges. Dashed lines show results from the original FLAMINGO L2p8\_m9 run, while solid lines show the effect of including 0.75 dex of scatter to the quasar luminosities. Adding scatter produces a broader ERDF that allows for super-Eddington ratios, and brings low-mass black holes to quasar-like bolometric luminosities.}

    \label{fig:eta}
\end{figure*}

In \cref{sec:results}, we showed that FLAMINGO successfully reproduces the number density of faint quasars with bolometric luminosities \(L_\mathrm{bol} \lesssim 10^{45}\,\ergs\). However, it underpredicts the observed number densities of more luminous quasars. To account for small-scale variability that is unresolved in our simulation, we introduced a log-normal scatter of 0.75 dex to the quasar luminosities. This boosts the abundance of bright quasars by up to 1 dex, bringing it closer to the observed values. Despite this improvement, FLAMINGO still fails to fully reproduce the observed rise and fall of the bright quasar population around cosmic noon. In this section, we aim to investigate what drives the differences between FLAMINGO and the observed bright end of the QLFs, and to study the consequences of adding scatter to our synthetic luminosities.

We first examine the Eddington ratio distribution function (ERDF) across different cosmic epochs. The Eddington ratio ($\lambda_{\rm Edd}$) is defined as the ratio of a quasar's bolometric luminosity ($L_\mathrm{bol}$) to its Eddington luminosity ($L_{\rm Edd}$):
\begin{equation}
\lambda_{\rm Edd} \equiv \frac{L_\mathrm{bol}}{L_{\rm Edd}}=\frac{L_{\rm bol}}{\epsilon_{\rm r}\dot{M}_{\mathrm{Edd}}c^2},
\end{equation}
where $\dot{M}_{\mathrm{Edd}}$ is the Eddington accretion rate defined in \cref{eq:edd_rate}.
\cref{fig:eta} shows the ERDF split in bins of black hole mass, with different colors representing different mass ranges. Dashed lines show the ERDF from the original FLAMINGO runs, while solid lines show the effect of adding scatter to the quasar luminosities. In general, we find that the Eddington ratios of quasars in the simulation span a very broad range, from $\approx10^{-5}$ to $\approx1$. In the original run, no quasar can exceed $\lambda_{\rm Edd}$ of unity, because black hole accretion is Eddington-limited (see Sec. \ref{sec:method}). When adding scatter, we are breaking this constraint by adding Gaussian noise to the quasars' Eddington ratios: as a result, super-Eddington values up to $\approx10$ (and, rarely, $\approx100$) are achieved at all redshifts. This has a direct implication: lower mass black holes ($M_\mathrm{BH}\lesssim10^8\,\msun$) are able to reach luminosities comparable to those of bright quasars. While this effectively inflates the number density of luminous quasars in the simulation, it lowers the average black hole mass of quasars at any given luminosity.

\begin{figure*}
    \centering
    \includegraphics[width=1.0\textwidth]{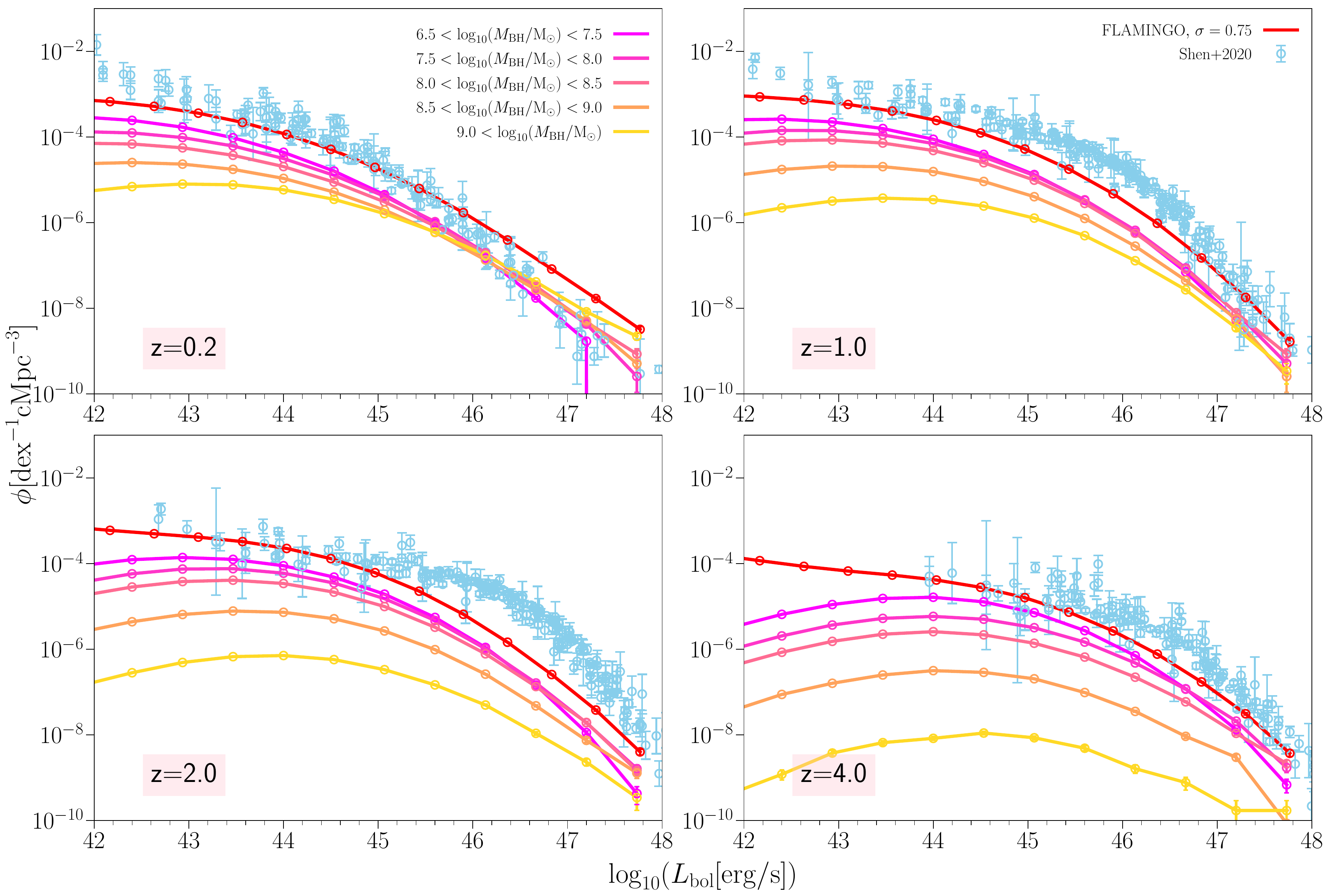}

    \caption{Decomposition of the quasar luminosity function in bins of black hole mass, for the case with added scatter. Different colors represent the contribution to the QLF of different black hole mass bins. The red lines show the total QLF from the simulation, and blue points are observational data from \citet{shen2020} (same as in \cref{fig:QLF}). Quasars with $M_\mathrm{BH} \sim 10^7$--$10^9\,\msun$ contribute comparably to the QLF across a wide luminosity range, while black holes with $M_\mathrm{BH} > 10^9\,\msun$ play a subdominant role except at low redshift ($z\lesssim1$) and very high luminosity ($L_\mathrm{bol} \gtrsim 10^{47}\,\ergs$).}
    \label{fig:decomp_mbh}
\end{figure*}

This effect can be observed in \cref{fig:decomp_mbh}, which shows the relative contribution to the QLFs of black holes with different masses, for the case with added scatter. The red lines and the observational data from \citet{shen2020} are the same as in \cref{fig:QLF}. \cref{fig:decomp_mbh} shows that, in general, a wide range of black hole masses ($10^7-10^9\,\msun$) contributes more or less equally to the QLF, for most quasar-like luminosities. Interestingly, the contribution of the most massive black holes ($M_\mathrm{BH}>10^9\,\msun$) is largely subdominant at all redshifts and luminosities, with the only exception of $z\lesssim1$ and $L_\mathrm{bol}\gtrsim10^{47}\,\ergs$.
This is in contrast with what is usually found in observations: luminous quasars ($L_\mathrm{bol}\gtrsim10^{47}\,\ergs$) have inferred black hole masses larger than $10^9\,\msun$ at all cosmic epochs, while fainter quasars $L_\mathrm{bol}\gtrsim10^{45.5}\,\ergs$ have black hole masses that are, on average, 1-1.5 dex lower. This is consistent with a very mild evolution of the measured Eddington ratios with bolometric luminosity, with most quasars showing ERDFs peaking in the range 0.1-0.5 \citep[e.g.,][]{wu_shen2022}.

Therefore, while adding scatter to quasar luminosities can partly reconcile the number density of bright quasars in the simulation with the observed one, it does so by including in the bright quasar population a large number of low-mass black holes emitting above the Eddington limit, which do not reflect the kind of black hole masses and Eddington ratios that are routinely observed. What is the reason for this mismatch? A plausible answer comes from \cref{fig:eta}: the shapes of the ERDFs for different black hole mass ranges are strikingly different at all redshifts. In particular, the accretion rate of the massive black hole population ($M_\mathrm{BH}\gtrsim10^9\,\msun$) is always strongly quenched ($\lambda_{\rm Edd} \approx 10^{-2}-10^{-3}$), while less massive black holes transition from highly accreting to dormant across cosmic time. Adding scatter to the luminosity of quasars does not alter this picture significantly, as it simply broadens the distributions of Eddington rates for a given mass and redshift. This has only a very minor effect for high-mass SMBHs, whose ERDF is already very broad and log-normal-like, but it severely impacts the shape of the low-mass black hole ERDFs, which are very narrow and peak at the Eddington limit. 

We conclude that the lack of bright quasars in FLAMINGO is not simply an issue of small-scale variability. It is a consequence of the fact that the most massive SMBHs do not make up the bright quasar population, in contrast to what is suggested by observations. The reason for this is likely twofold: very massive black holes may be too rare in FLAMINGO -- possibly because the high-redshift growth of SMBHs is not rapid enough -- and their accretion may be hindered by feedback mechanisms or by the lack of sufficient fueling material.   
By integrating the ERDF distribution for  $M_\mathrm{BH}\gtrsim10^9\,\msun$ SMBHs at $z\approx4$ in \cref{fig:eta}, we find that the predicted number density of these SMBHs in FLAMINGO is $\approx10^{-7}-10^{-8}\,\cMpc^{-3}$. This is broadly compatible with the constraints on the black hole mass function (BHMF) of e.g., \citet{he2024} at similar redshifts. This implies that, in order to respect observational constraints, essentially all high-$z$ massive black holes in FLAMINGO should be active, in contrast with what is found in \cref{fig:eta}.

\subsection{The quasar-dark matter halo connection across cosmic time} \label{sec:discussion_clustering}

\begin{figure*}
    \centering
    \includegraphics[width=1.0\textwidth]{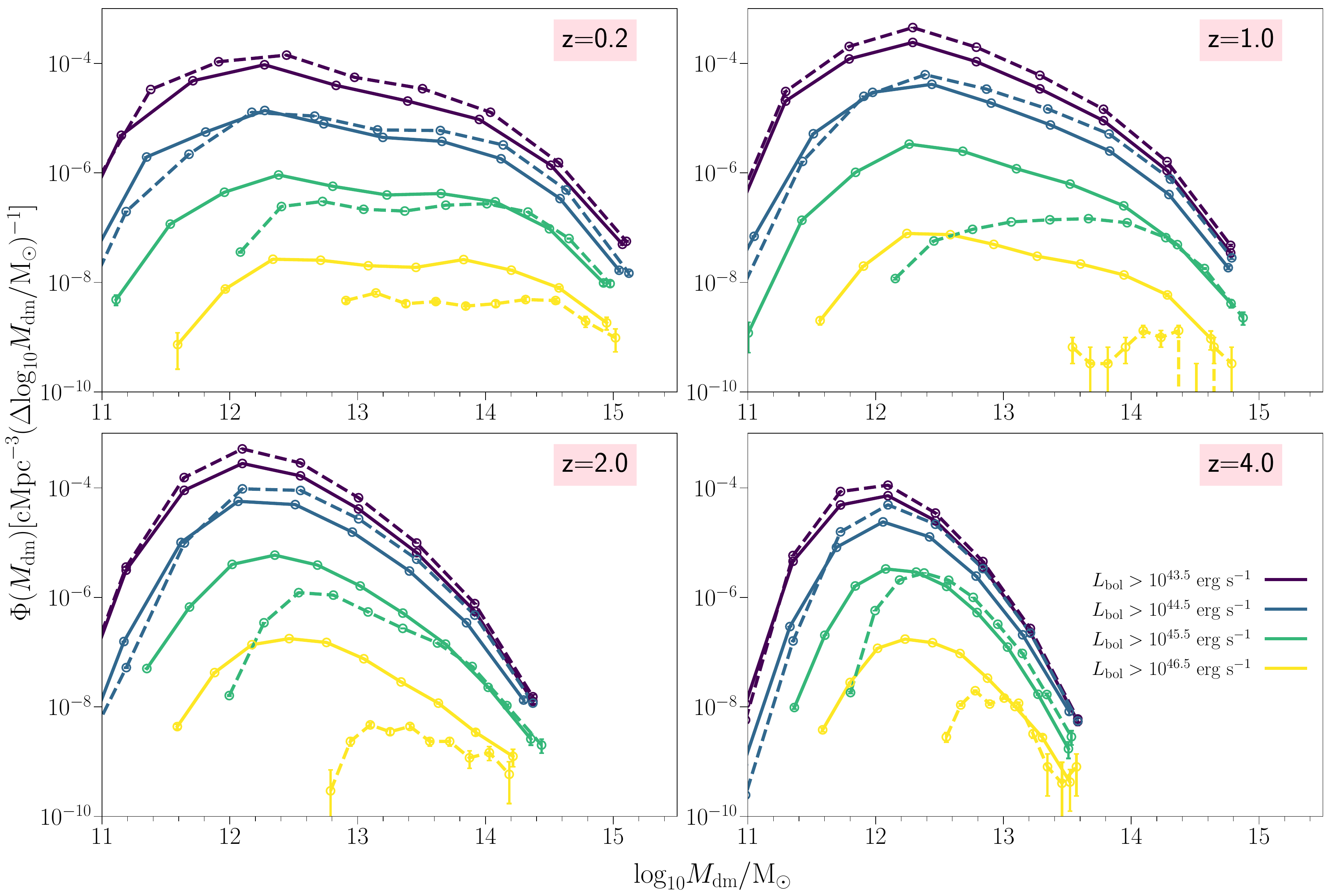} 

    \caption{Quasar-host mass functions (QHMFs; i.e., mass function of quasar-hosting haloes) for different bolometric luminosity thresholds (colored lines). Dashed lines show the QHMFs for the original FLAMINGO L2p8\_m9 run, while solid lines show the effects of including 0.75 dex of scatter in quasar luminosity. Adding scatter broadens the distribution and, for the brightest quasars, shifts the peak of the QHMFs toward lower halo masses. Here, halo masses are defined according to the $M_{\rm 200c}$ definition (spherical overdensity mass enclosing 200 times the critical density), and only central haloes are included.}

    \label{fig:qhmf}
\end{figure*}

In cosmological simulations, quasars are hosted by dark matter haloes, and their spatial clustering is tightly linked to the clustering properties of these haloes. Since dark matter haloes are the building blocks of large-scale structure, understanding quasar clustering provides valuable insight into the connection between luminous active galactic nuclei and the underlying matter distribution. To better understand the origin of the predicted clustering of quasars in FLAMINGO, we show in \cref{fig:qhmf} the quasar-host mass function (QHMF) for different quasar luminosity thresholds $L_\mathrm{bol, thr}$ -- i.e., the mass distribution of haloes hosting quasars brighter than $L_\mathrm{bol, thr}$. Once again, we show the results coming directly from the FLAMINGO L2p8\_m9 run (dashed line) and those obtained by adding scatter to the luminosity of quasars (solid). In this analysis we include only central haloes, and adopt the $M_{\rm 200c}$ spherical overdensity definition for the halo mass. 

The QHMFs in \cref{fig:qhmf} show a relatively consistent evolution with both redshift and luminosity. At high $z$, the distributions show a clear peak in halo mass and a relatively narrow window in possible halo hosts, with a definite log-normal shape. Beyond cosmic noon, the QHMFs become much broader for all luminosity thresholds, with the $z=0.2$ case showing a large range of candidate quasar hosts and skewed as well as possibly multimodal distributions.

The luminosity dependence of the QHMFs depends strongly on whether scatter is added to the simulated quasar luminosities. In particular, for high luminosity thresholds, the shapes of the QHMFs differ markedly between the ``with'' and ``without'' scatter cases, as is evident from \cref{fig:qhmf}. This difference arises because the Eddington bias becomes increasingly important toward the bright end, promoting lower-mass haloes into the luminous quasar population when scatter is included. Moreover, for the case without any additional scatter, the peak of the QHMF (which primarily determines the clustering strength of quasars) is a highly non-linear function of the luminosity threshold $L_\mathrm{bol, thr}$. In particular, lower thresholds $L_\mathrm{bol, thr}\lesssim10^{45}\,\ergs$ correspond to similar QHMF distributions broadly peaking at $M_\mathrm{halo}\approx10^{12}-10^{12.5}\,\msun$, in general agreement with observational constraints \citep[e.g.,][]{arita2023}. For larger luminosity thresholds, however, the peak rapidly shifts to much larger halo masses, reaching values as high as $M_\mathrm{halo}\approx10^{13}-10^{13.5}\,\msun$ for the brightest threshold considered ($L_\mathrm{bol, thr}=10^{46.5}\,\ergs$). This implies a rapid increase in clustering with luminosity for the bright quasar population. Interestingly, such rapid evolution would imply a strong clustering of bright quasars, in line with that measured by \citet{Shen_2006} at $z\approx4$. Indeed, the QHMF we find for the brightest luminosity threshold at $z\approx4$ is similar to that derived in \citet{pizzati2024a} by matching the \citet{Shen_2006} constraints. \citet{pizzati2025} also predict a significant dependence of clustering on luminosity at high $z$ based on their modeling of the \citet{Shen_2006} observations: their faint-quasar QHMF agrees well with our findings for the same redshift and $L_\mathrm{bol, thr}$ (dashed green line). However, we caution that the number of bright quasars in the case with no added scatter is much lower in FLAMINGO than what observations predict\footnote{For this reason, it was not possible to compute a clustering signal for the bright quasar population in the no-scatter case, and the only predictions reported in Fig.~\ref{fig:clustering} are obtained by setting a lower luminosity threshold to include the more abundant fainter quasar population.}, so the strong dependence of the QHMF on luminosity at the bright end may be driven by small samples of quasars living in highly biased environments, and not reflect the actual distribution of the entire sample of luminous quasars found in observations.

Indeed, when 0.75 dex of scatter are added to the simulated quasar luminosities (solid lines in \cref{fig:qhmf}), the luminosity dependence of quasar clustering is washed out almost completely. In this case, the resulting QHMFs consistently peak around $M_\mathrm{halo}\approx10^{12}\,\msun$ for essentially all luminosity thresholds and a wide range of redshifts. The very shallow (or non-existent) trend of clustering with luminosity is consistent with that observed by several studies \citep[e.g.,][]{adelberger_steidel2005, Shen_2009, Eftekharzadeh_2015}, who find insignificant differences in the clustering of bright and faint quasars at $0\lesssim z\lesssim3$. The $M_\mathrm{halo}\approx10^{12}\,\msun$ value for typical quasar hosts is also predicted by observations as well as theoretical models. haloes of this mass are the most efficient at converting gas into stars and at fuelling central black holes \citep[e.g.,][]{dekel09,wechsler18}. Below this scale, supernova–driven outflows can expel gas and suppress both star formation and black‐hole growth \citep{Bower2017, Dubois_2015}.
Above it -- particularly beyond \(10^{13}\,\msun\) -- infalling gas is shock-heated and forms a quasi-static hot halo \citep[e.g.,][]{correa2018, dekel09}; the resulting cooling time exceeds the dynamical time and prevents gas condensation and star formation \citep{rees77}.
Consequently, haloes in the \(10^{12}\)–\(10^{13}\ \msun\) range strike the optimal balance between being massive enough to host luminous quasars while still permitting efficient gas cooling -- consistent with the ``sweet spot'' we find for the QHMFs in \cref{fig:qhmf}.

While the luminosity dependence found for the QHMFs in the case with added scatter is encouraging -- at least for $z\lesssim3$ -- we note that, by adding scatter, we are effectively lowering the average SMBH mass for any fixed quasar luminosity. In \cref{sec:discussion_qlf}, we have argued that the resulting SMBH masses are too low to be compatible with those observed for bright quasars by surveys such as the SDSS \citep{wu_shen2022}. Given the well-established correlation between SMBH mass and the mass of their host dark matter haloes \citep[e.g.,][]{ferrarese2002,Booth2010}, attributing bright quasars to lower mass SMBHs leads to lowered typical quasar host halo masses, since lower mass SMBHs tend to live in lower mass haloes. We therefore interpret the two cases considered in our analysis -- those “with” and “without” added scatter in quasar luminosity -- as bracketing the plausible range of outcomes for the QHMF. These two scenarios effectively define limiting behaviors for how quasar clustering depends on luminosity and redshift, capturing the sensitivity of our predictions to uncertainties in the underlying black hole-halo connection. More detailed models and refined quasar clustering observations will be able to pinpoint this behavior even further, especially in the high-redshift regime.

\section{Conclusions}
\label{sec:conclusions}

In this paper, we use the \textsc{FLAMINGO} suite of cosmological hydrodynamical simulations to study the properties of the bright quasar population. We focus primarily on the L2p8\_m9 simulation, which offers an extremely large box of $L=2.8\,\cGpc$ run at \textsc{FLAMINGO}'s intermediate resolution. Thanks to the large volume probed by the simulation, we aim to characterize the properties of accreting SMBHs by focusing on two key observables: the quasar luminosity function (QLF), and the clustering (auto-correlation function) of quasars.

Our analysis reveals that the QLF in FLAMINGO reproduces observational constraints at low redshift and for faint quasars ($L_\mathrm{bol} \lesssim 10^{45}\,\mathrm{erg\,s^{-1}}$). However, the simulation significantly underestimates the abundance of bright quasars, particularly around cosmic noon ($z\approx1$--3; \cref{fig:QLF}). We experiment with introducing a log-normal scatter of 0.75 dex to the quasars' bolometric luminosities, in order to account for the small-scale variability that is not resolved in the simulation. This also implies that the quasars can shine above the Eddington limit, which is not allowed in the original FLAMINGO model. Adding scatter partially bridges the gap at the bright end, improving the match without compromising the faint end. Still, substantial discrepancies remain, pointing to limitations in the subgrid prescriptions for SMBH accretion and feedback and underscoring the need for higher numerical resolution. 

In particular, by studying the Eddington ratio distribution function (ERDF; \cref{fig:eta}) and by decomposing the QLF in bins of black hole mass (\cref{fig:decomp_mbh}), we find that the mismatch between the simulation and observations at the bright end is mainly driven by two key issues: massive black holes ($M_\mathrm{BH}\gtrsim10^9\,\msun$) in the simulation are too rare -- likely due to insufficient early growth -- and their accretion is strongly quenched at all redshifts ($\lambda_\mathrm{Edd} \sim 10^{-2}$–$10^{-3}$), making them unable to account for the bright quasar population. While adding scatter to the quasar luminosities produces many more bright quasars, these are mainly lower-mass SMBHs with super-Eddington accretion rates. SMBH mass measurements from quasar spectra \citep[e.g.,][]{wu_shen2022}, instead, infer bright quasars to have large black hole masses resulting in moderate Eddington ratios in the range $\lambda_\mathrm{Edd} \sim 0.1-1$.

In addition to studying the QLF, we analyzed the clustering properties of quasars using the projected auto-correlation function as a clustering statistic (\cref{fig:clustering}). The FLAMINGO simulation reproduces quasar clustering reasonably well at $z \lesssim 3$, especially around $z \approx 2.5$ where tight constraints from the BOSS survey are available \citep[][]{Eftekharzadeh_2015}. At \(z \approx 4\) the simulation predicts significantly weaker clustering than observed by \citet{Shen_2006}, reflecting a long-standing tension between high-$z$ clustering measurements and theoretical models and simulations \citep[e.g.,][]{white_2008, shankar_2010}. We find that adding scatter to the luminosity of quasars reduces the overall clustering strength. This effect arises because scatter increases the luminosity of quasars in lower-mass haloes, which are more numerous and less strongly clustered. As a result, the contribution from less massive haloes dilutes the overall clustering signal. 

We show that the correlation length increases with both redshift and luminosity (\cref{fig:c_length}). Without additional scatter, clustering shows a clear dependence on luminosity, which is even more apparent by looking at the quasar-host mass functions (QHMFs; \cref{fig:qhmf}). After adding scatter, this dependence becomes weaker, since more low-mass haloes contribute to the bright quasar sample. The presence of significant scatter in the luminosity-halo mass relation is indeed what likely drives the little to no luminosity dependence of quasar clustering found in observations \citep{adelberger_steidel2005,Shen_2009, Eftekharzadeh_2015}.

To better understand the relation between quasar and host haloes in FLAMINGO, we analyze the quasar-host mass function (QHMF) across redshift and luminosity (\cref{fig:qhmf}). We find that more luminous quasars reside in more massive haloes, especially at high redshift, where the QHMF peaks around $M_\mathrm{halo} \approx 10^{12}$--$10^{13}\,\msun$, consistent with expectations from halo-based models. These halo masses strike a balance between efficient gas cooling and sufficient gravitational potential to host luminous quasars. Our QHMF and clustering analyses emphasize the central role of $\sim10^{12}$--$10^{13}\,\msun$ haloes in shaping the spatial distribution of quasars. They also highlight the need to carefully account for luminosity scatter when interpreting quasar clustering and its connection to halo occupation.

Further understanding of quasars and their cosmic evolution will require both improved theoretical models and tighter observational constraints. Our findings suggest that capturing the full diversity of quasar activity -- particularly at the bright end -- depends on simulations with better resolution and more realistic treatments of SMBH growth and feedback, including mechanisms that allow rapid early accretion. At the same time, better measurements of the quasar luminosity function and clustering, especially at high redshift, are essential to test and refine simulation predictions. This will be possible with current and upcoming surveys such as DESI \citep{desicollaboration2016} and Roman \citep{spergel2015}.

\section*{Acknowledgements}

We are grateful to the FLAMINGO team for making their simulations available. 
We thank Rob McGibbon for help with the simulation outputs.
We acknowledge helpful conversations with the ENIGMA group at
UC Santa Barbara and Leiden University. 
JFH and EP acknowledge support from the European Research Council (ERC) under the European
Union’s Horizon 2020 research and innovation program (grant agreement No 885301).
WM acknowledges support by the Dutch Research Council (NWO) through the Dark Universe Science Collaboration (OCENW.XL21.XL21.025).
This work used the DiRAC@Durham facility managed by the Institute for
Computational Cosmology on behalf of the STFC DiRAC HPC Facility
(\url{www.dirac.ac.uk}). The equipment was funded by BEIS capital funding via
STFC capital grants ST/K00042X/1, ST/P002293/1, ST/R002371/1 and ST/S002502/1,
Durham University and STFC operations grant ST/R000832/1. DiRAC is part of the
National e-Infrastructure.

\section*{Data Availability}

The derived data generated in this research will be shared on reasonable requests to the corresponding author.



\bibliographystyle{mnras}
\bibliography{biblio} 


\appendix

\section{Convergence tests} \label{appendix:convergence}

\begin{figure*}
    \centering
    \includegraphics[width=1.0\textwidth]{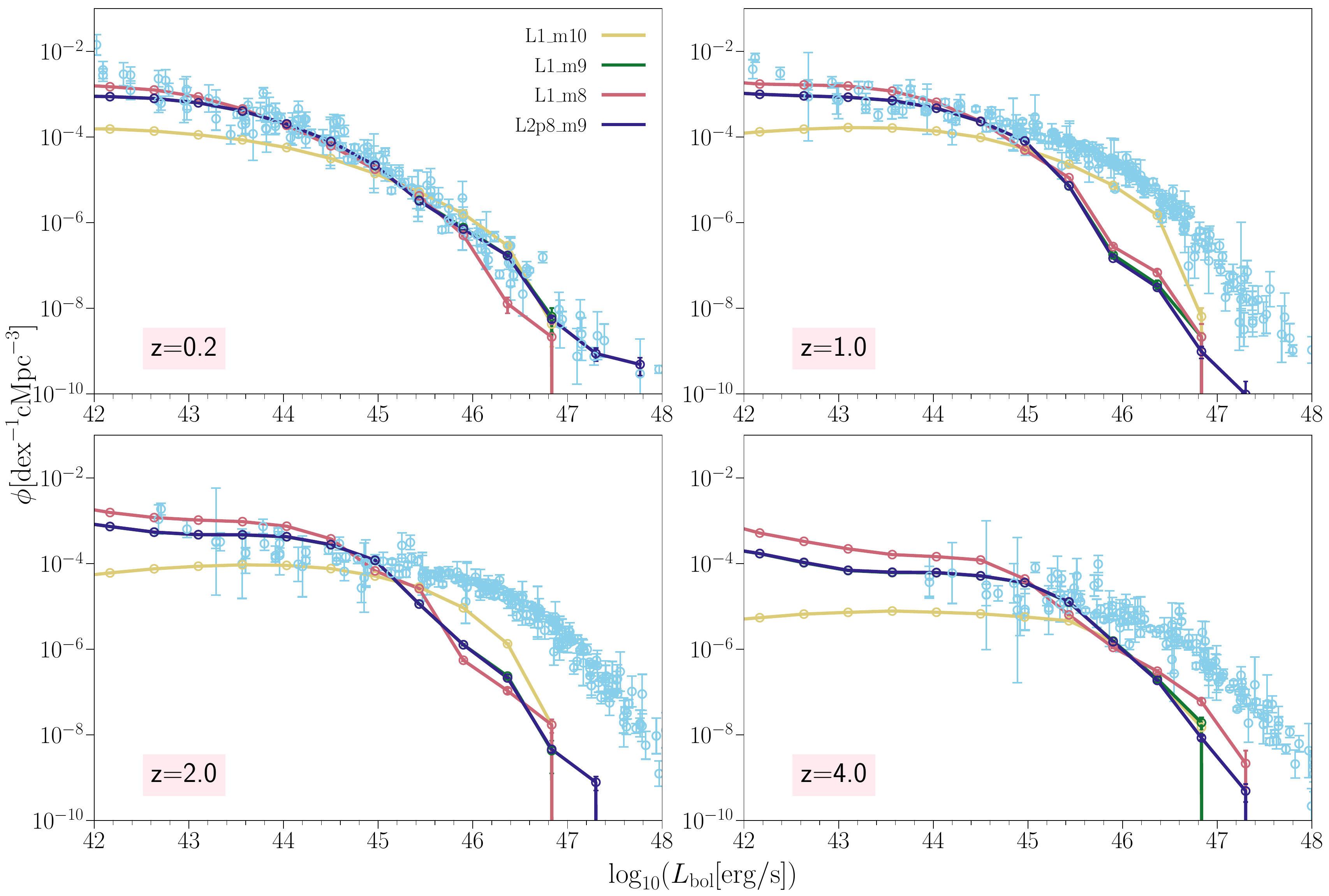} 
    
    \caption{QLFs from simulations with different box sizes and resolutions. All lines represent simulation results, while blue points denote observational data (see \cref{fig:clustering}). $\mathrm{L1\_m10}$, $\mathrm{L1\_m9}$, and $\mathrm{L1\_m8}$ correspond to $1\ \mathrm{Gpc}$ boxes with low, intermediate, and high resolution, respectively. $\mathrm{L2p8\_m9}$ represents a $2.8\,\mathrm{Gpc}$ box with intermediate resolution. The results show good convergence at the bright end, while differences at the faint end are primarily resolution-driven.}
    \label{fig:reso}
\end{figure*}

In this section, we perform a convergence test to examine the impact of simulation box size and resolution on the results. The relevant simulation parameters are summarized in \cref{tab:runs}. As indicated by the naming convention, $\mathrm{L1\_m10}$, $\mathrm{L1\_m9}$, and $\mathrm{L1\_m8}$ refer to simulations with a $1\ \mathrm{Gpc}$ box and low, intermediate, and high resolution respectively, while $\mathrm{L2p8\_m9}$ corresponds to a $2.8\ \mathrm{Gpc}$ box with intermediate resolution. \cref{fig:reso} presents the resulting QLFs from these simulations. Colored lines represent different runs, and blue points show observational data (same as \cref{fig:QLF}). We find that the results are converged with respect to box size in the range \(10^{42}\!-\!10^{46}\,\mathrm{erg\,s^{-1}}\), while at the bright end (\(L_{\mathrm{bol}} \gtrsim 10^{46}\,\mathrm{erg\,s^{-1}}\)) the box size starts to have a slight impact. Overall, the differences between simulations with different box sizes remain small, and the larger volume, as expected, enables us to probe quasars at lower number densities and higher luminosities.

In contrast, the impact of resolution is more significant, especially at the low-luminosity end of the QLF. The $\mathrm{L1\_m10}$ simulation with low resolution shows substantial deviations from the other runs across all redshifts. The intermediate-resolution ($\mathrm{L1\_m9}$) and high-resolution ($\mathrm{L1\_m8}$) simulations agree well at the bright end, indicating successful convergence, but their differences become increasingly apparent for high redshifts and low luminosities. However, in this regime, we lack sufficient observational data for direct comparison. Within the luminosity range covered by the observations, the intermediate-resolution and high-resolution results are very similar, indicating good convergence.

\begin{figure*}
    \centering
    \includegraphics[width=1.0\textwidth]{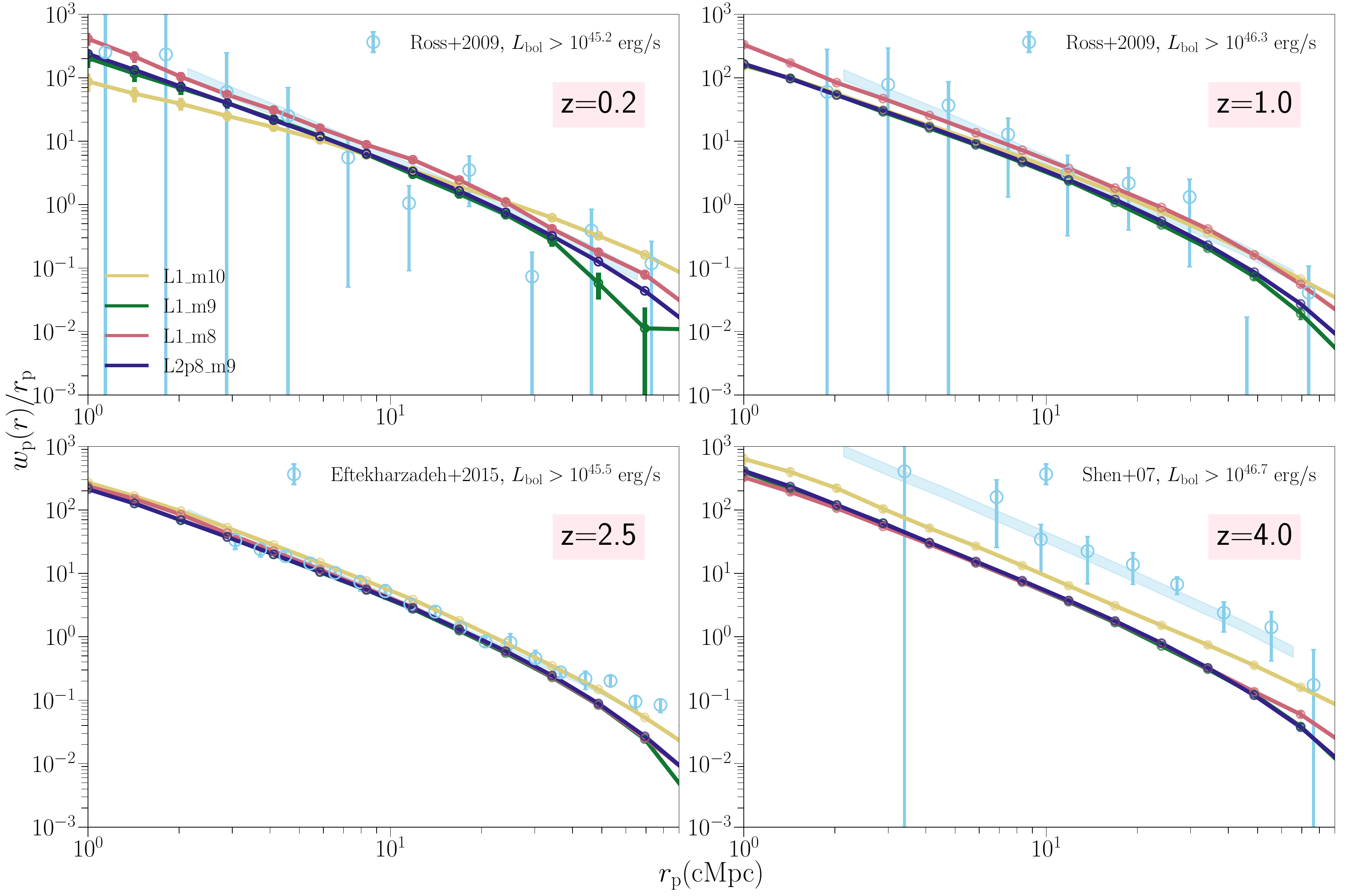} 
    
    \caption{Projected correlation function from simulations with different box sizes and resolutions compared to observations. Colored lines represent different simulations. Blue points are from the observations (see \cref{fig:clustering}). The luminosity threshold for quasars in the simulations is set to \(L_{\mathrm{bol,thr}}=10^{45}\ \mathrm{erg\ s^{-1}}\).}
    \label{fig:reso_clustering}
\end{figure*}

In \cref{fig:reso_clustering}, we investigate the impact of box size and resolution on quasar clustering, using a luminosity threshold of \(L_{\mathrm{bol,thr}}=10^{45}\ \mathrm{erg\ s^{-1}}\).We find that our clustering results are converged with respect to the box size, with only a slight impact visible on large scales (\(r \gtrsim 30\,\mathrm{cMpc}\)). However, the larger simulation volume provides a significantly larger quasar sample, which leads to smaller error bars and enables more robust comparisons with observational data. 
Different resolutions show generally good convergence, with the $\mathrm{L1\_m9}$ and  $\mathrm{L1\_m8}$ runs being much more similar than $\mathrm{L1\_m10}$. However, at $z=1.0$ (and partly $z=0.2$) the high-resolution run predicts stronger clustering than the intermediate-resolution simulations. The reason for this difference is unclear and suggests that our results may not be entirely converged at the lowest redshifts. The large observational uncertainties at these epochs, however, are still more important than the deviations between the different runs.

\section{Results for model variations}
\label{appendix:variations}

\begin{figure*}
    \centering
    \includegraphics[width=1.0\textwidth]{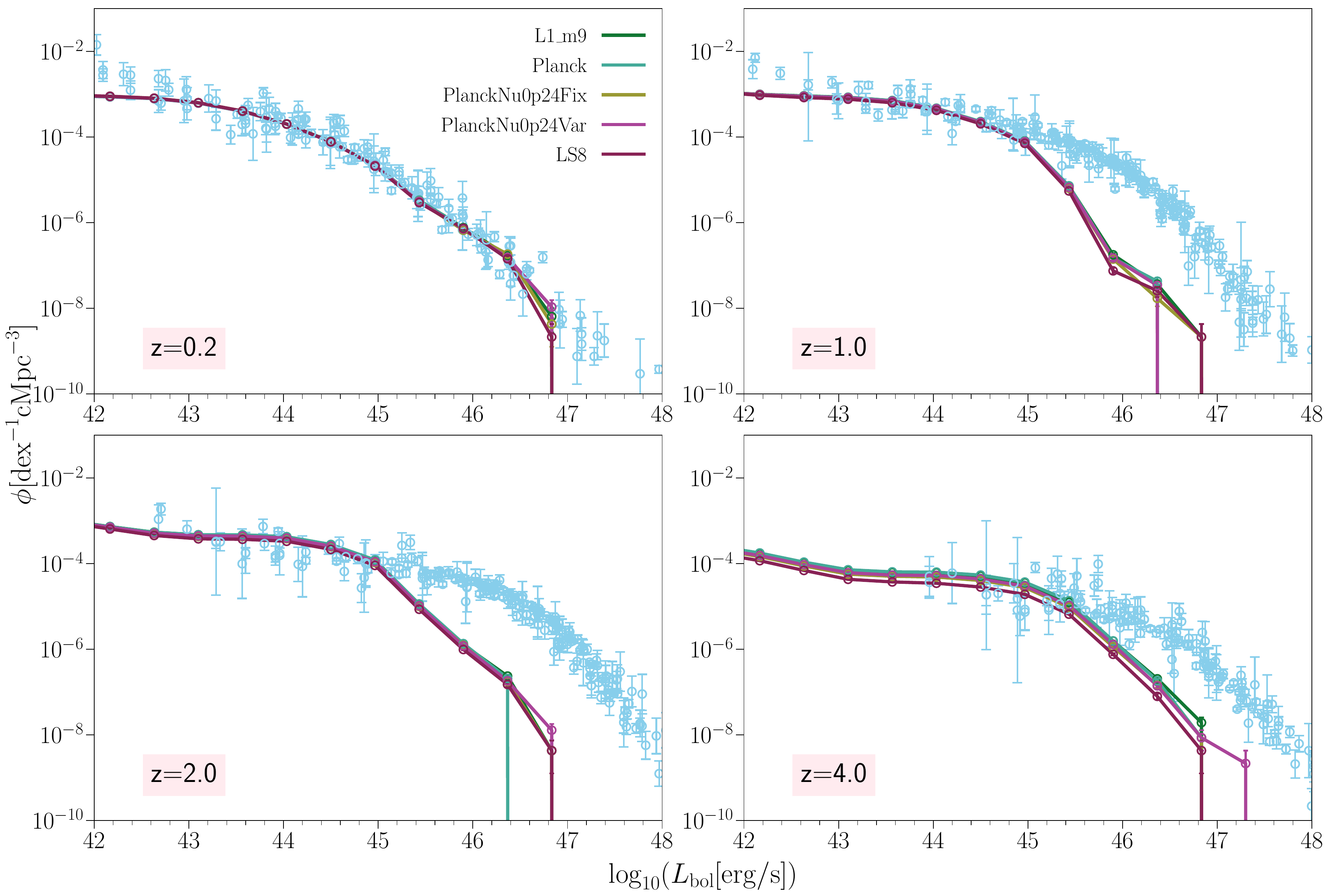} 
    
    \caption{QLFs from simulations with different cosmological variations in a $1\ \mathrm{Gpc}$ box. Different colors indicate different cosmologies, including variations in the total neutrino mass and the amplitude of the matter power spectrum. Blue points show the observational data compiled by \citet{shen2020} (same as \cref{fig:QLF}). 
    }
    \label{fig:cos}
\end{figure*}

Besides the fiducial cosmology, there are four cosmological variations for a spatially flat universe in FLAMINGO. These variations include the Planck cosmology \citep{Planck_2020}, one with the total neutrino mass \(\Sigma\ m_\nu=0.06\ \mathrm{eV}\), two with \(\Sigma\ m_\nu=0.24\ \mathrm{eV}\), and one with a suppressed amplitude of the matter power spectrum, which is preferred by the large-scale structure survey \citep{Amon_2023}. The cosmological parameters for each different run are listed in \cref{tab:numerical_parameters}. All the variations use the same calibration data and are run in a $1\ \mathrm{Gpc}$ box size. Here and in Appendix \ref{appendix:convergence}, we focus on the original FLAMINGO runs and do not add any scatter to quasar luminosities.
\begin{table*}
	\centering
	\caption{Cosmological parameters for different simulations. The values from left to right are: The dimensionless Hubble constant $h$; the total matter density parameter $\Omega_\mathrm{m}$; the dark energy density parameter $\Omega_\Lambda$; the total baryonic matter density parameter $\Omega_\mathrm{b}$; the combined particle mass of all neutrino species $\sum m_\nu$; the amplitude of the primordial matter power spectrum $A_s$,  the primordial matter power spectral index $n_s$; the amplitude of the initial power spectrum, defined as the root-mean-square mass density fluctuation within spheres of radius $8 h^{-1}\mathrm{Mpc}$, extrapolated to $z=0$ using linear theory $\sigma_8$; the amplitude of the initial power spectrum parametrized as $S_8\equiv \sigma_8\sqrt{\Omega_m/0.3}$; the neutrino matter density parameter $\Omega_\gamma$. The table is from \citet{FLAMINGO}.}
	\label{tab:numerical_parameters}
\begin{tabular}{lcccccccccc}
\hline Prefix & $h$ & $\Omega_{\mathrm{m}}$ & $\Omega_{\Lambda}$ & $\Omega_{\mathrm{b}}$ & $\sum m_\nu c^2$ & $A_{\mathrm{s}}$ & $n_{\mathrm{s}}$ & $\sigma_8$ & $S_8$ & $\Omega_\nu$ \\
\hline- & $0.681$ & $0.306$ & $0.694$ & $0.0486$ & $0.06 \mathrm{eV}$ & $2.099 \times 10^{-9}$ & $0.967$ & $0.807$ & $0.815$ & $1.39 \times 10^{-3}$ \\
Planck & $0.673$ & $0.316$ & $0.684$ & $0.0494$ & $0.06 \mathrm{eV}$ & $2.101 \times 10^{-9}$ & $0.966$ & $0.812$ & $0.833$ & $1.42 \times 10^{-3}$ \\
PlanckNu0p12Var & $0.673$ & $0.316$ & $0.684$ & $0.0496$ & $0.12 \mathrm{eV}$ & $2.113 \times 10^{-9}$ & $0.967$ & $0.800$ & $0.821$ & $2.85 \times 10^{-3}$ \\
PlanckNu0p24Var & $0.662$ & $0.328$ & $0.672$ & $0.0510$ & $0.24 \mathrm{eV}$ & $2.109 \times 10^{-9}$ & $0.968$ & $0.772$ & $0.807$ & $5.87 \times 10^{-3}$ \\
PlanckNu0p24Fix & $0.673$ & $0.316$ & $0.684$ & $0.0494$ & $0.24 \mathrm{eV}$ & $2.101 \times 10^{-9}$ & $0.966$ & $0.769$ & $0.789$ & $5.69 \times 10^{-3}$ \\
LS8 & $0.682$ & $0.305$ & $0.695$ & $0.0473$ & $0.06 \mathrm{eV}$ & $1.836 \times 10^{-9}$ & $0.965$ & $0.760$ & $0.766$ & $1.39 \times 10^{-3}$ \\
\hline
\end{tabular}
\end{table*}
In \cref{fig:cos}, we compare the effects of different cosmological variations on the results of the QLF. Different colors represent different cosmological variations, and the blue points show the observational data from \citet{shen2020} (see also \cref{fig:QLF}). We find that the effect of different cosmological variations is minimal. The only exception is that at high redshifts ($z > 1$), the QLF predicted by the low $\sigma_8$ (LS8) run is systematically lower than those from the other cosmological variations. 
This suppression likely arises because a smaller $\sigma_8$ value implies weaker matter clustering and thus fewer massive haloes. Since luminous quasars preferentially reside in relatively massive haloes (see, e.g., \cref{fig:qhmf}), the reduced halo abundance leads to a lower number density of bright quasars and hence a lower QLF.

\begin{figure*}
    \centering
    \includegraphics[width=1.0\textwidth]{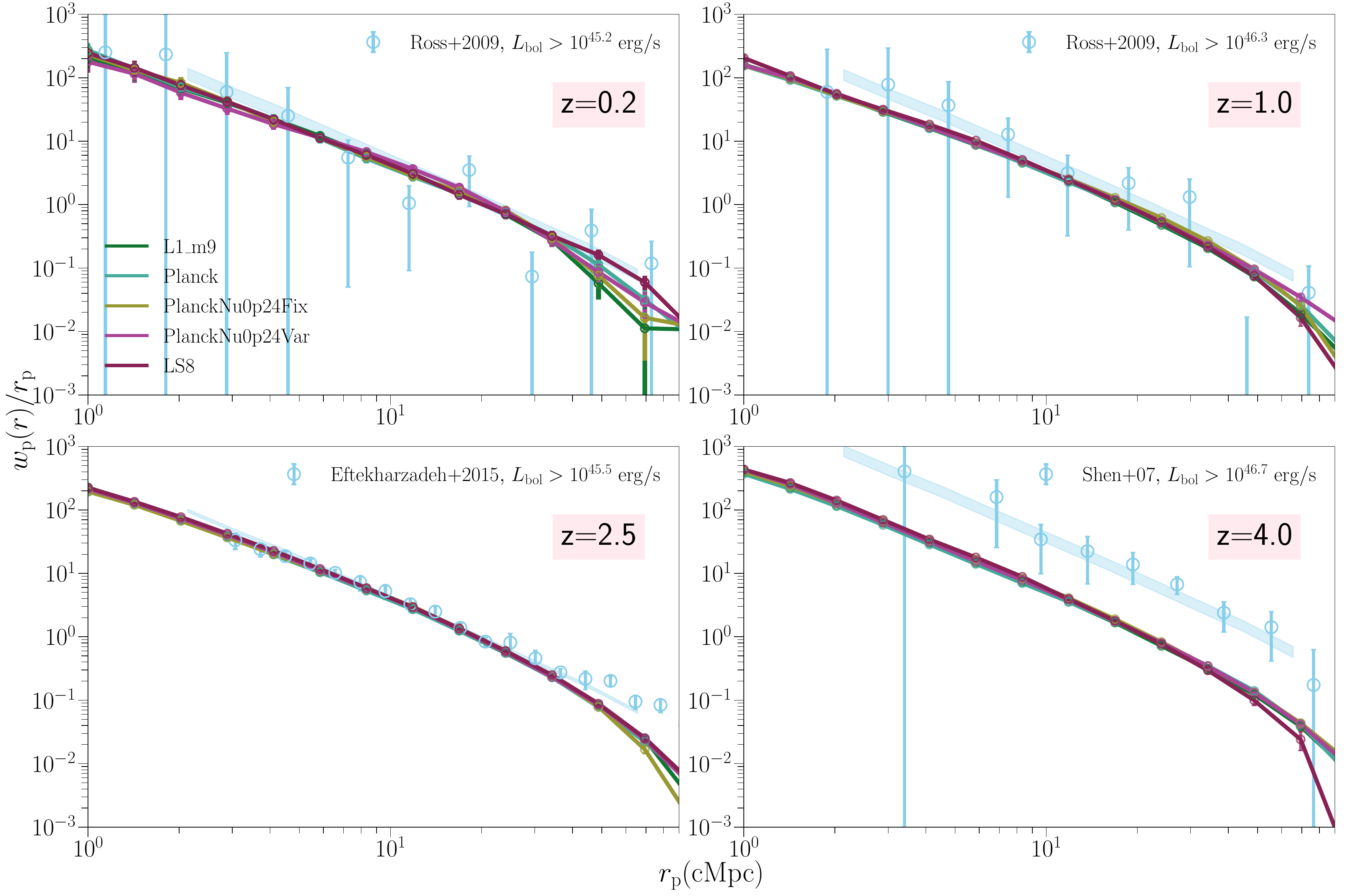} 
    
    \caption{Projected two-point correlation functions of quasars from simulations with different cosmological models. Different colors represent different cosmological variations, and blue points are the observational data (see \cref{fig:clustering}). The luminosity threshold for quasars in the simulations is \(L_{\mathrm{bol,thr}}=10^{45}\ \mathrm{erg\ s^{-1}}\).}
    \label{fig:cos_clustering}
\end{figure*}

In \cref{fig:cos_clustering}, we present the impact of cosmology variations on quasar clustering. Different colors represent different cosmological models, while blue points indicate the observational data shown in \cref{fig:clustering}. 
The luminosity threshold is set to \(L_{\mathrm{bol,thr}}=10^{45}\ \mathrm{erg\ s^{-1}}\) for all redshifts. Note that this choice is different than the one we made in \cref{fig:clustering}, where we were aiming to directly reproduce observations. Here, observations only serve as a reference, and we employ a lower luminosity threshold due to the smaller simulation volume ($L=1\,\cGpc$) than that used in the main text ($L=2.8\,\cGpc$). A smaller box size yields fewer luminous quasars, and applying a higher threshold would result in poor statistics and increased noise in the clustering measurement. Overall, we find that cosmological variations have a relatively minor effect on the clustering strength across all redshifts. The clustering amplitudes are nearly identical among all cosmological models, including the low $\sigma_8$ case. This indicates that the reduced number of quasars in the low $\sigma_8$ cosmology does not translate into a noticeable difference in clustering strength.

In addition to the fiducial simulation, \textsc{FLAMINGO} also includes eight astrophysical variations. Here, fgas refers to runs in which the observed cluster gas fractions are systematically shifted by $\pm N\sigma$ relative to their measurement uncertainties, while M$^*$ refers to runs in which the observed stellar masses are systematically reduced by 0.14 dex. The Jet runs adopt a kinetic, jet-like mode of AGN feedback instead of the fiducial thermal implementation. The specific subgrid parameters for these models are detailed in \citet{FLAMINGO}.

\begin{figure*}
    \centering
    \includegraphics[width=1.0\textwidth]{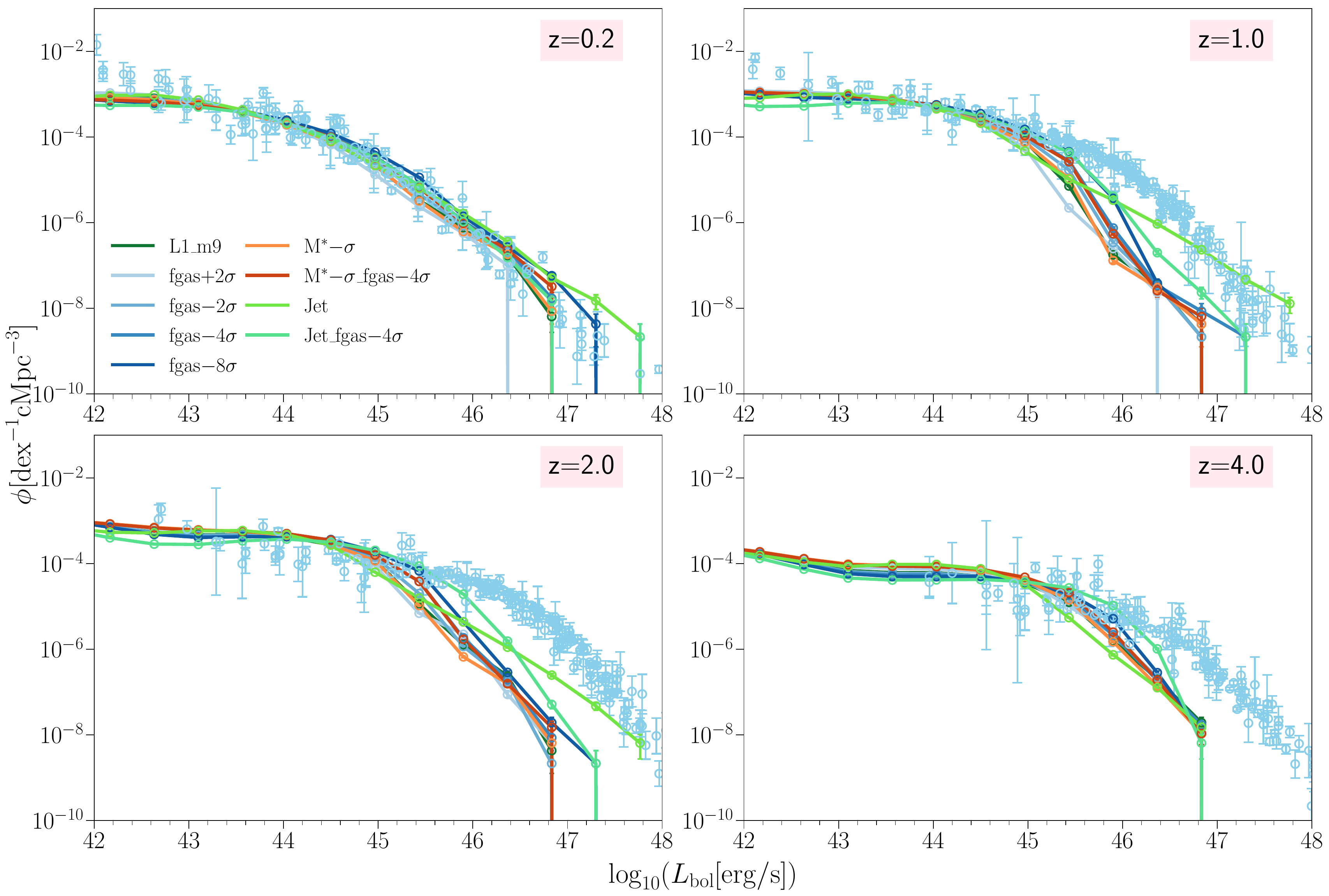} 
    
    \caption{QLFs from simulations with different astrophysical models in a $1\ \mathrm{Gpc}$ box. Different colors represent various feedback models, and blue points are the observational data (see \cref{fig:QLF}). The differences among models are minor at the faint end, while simulations with jet feedback better reproduce the observed QLF at the bright end.}
    \label{fig:phy}
\end{figure*}

All these simulations have a box size of $1\ \mathrm{Gpc}$. In \cref{fig:phy}, we show the results of the QLF with different astrophysical variations. Different colors represent different astrophysical models, and the blue points are from the observation (same as \cref{fig:QLF}).
We find that at the low-luminosity end, the differences between the various astrophysical models are not significant. At the bright end, the simulations with jet feedback show better agreement with the observational data. This improvement can be attributed to the distinct nature of kinetic jet feedback compared to isotropic thermal feedback. Jet feedback operates through large-scale kinetic outflows and acts on longer timescales, allowing AGN to remain luminous for extended periods. In contrast, isotropic thermal feedback in other models quenches AGN activity more rapidly, resulting in shorter-lived bright phases. Consequently, jet feedback statistically results in a higher number of bright AGN being active at any given redshift, improving the match with the observed QLF. We also note that all non-jet simulations produce very similar results, suggesting that variations in astrophysical parameters have a relatively minor impact on the final QLFs.

\begin{figure*}
    \centering
    \includegraphics[width=1.0\textwidth]{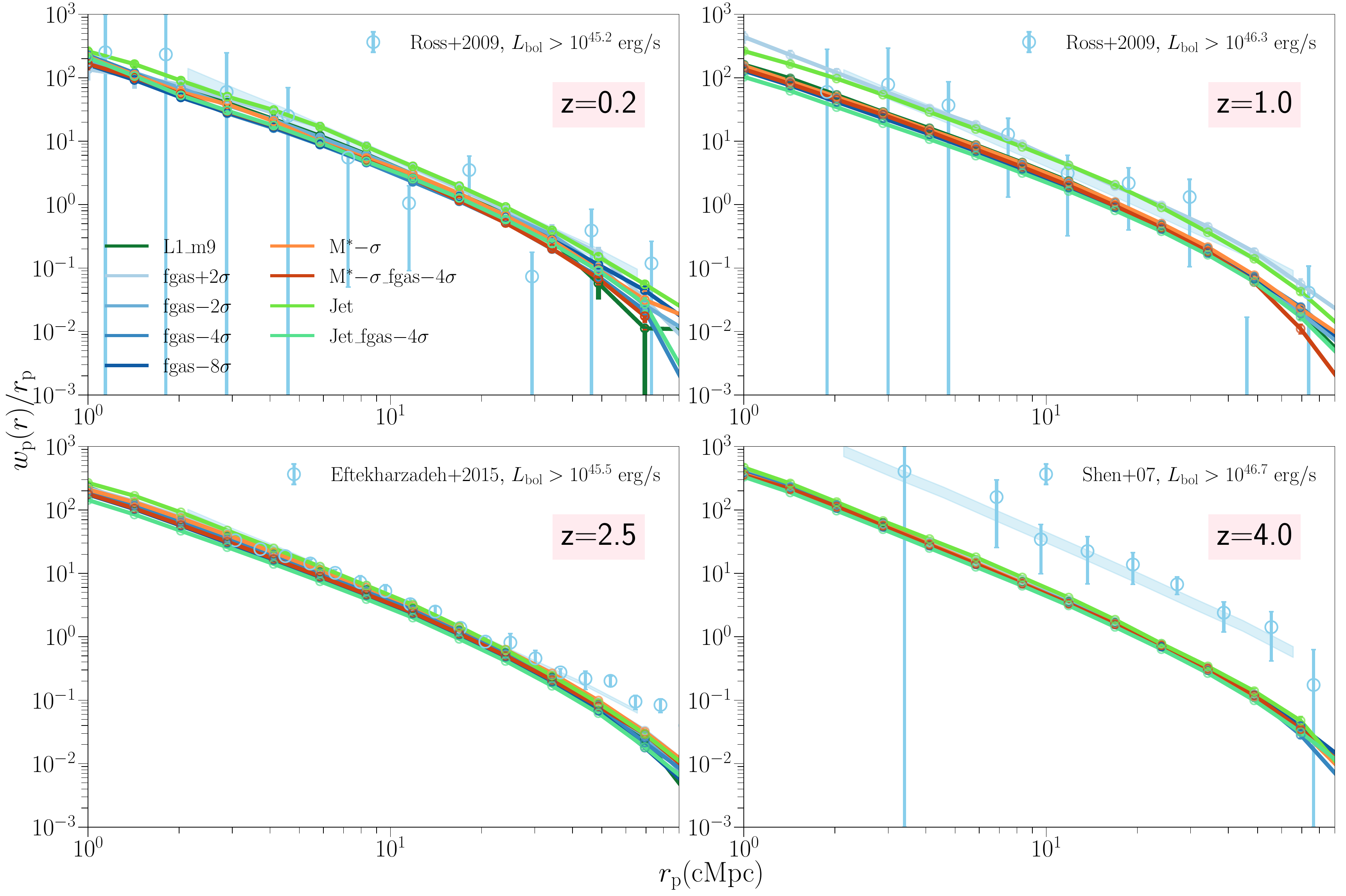} 
    
    \caption{Projected two-point correlation function for quasars from simulations with different astrophysical feedback models in a $1\ \mathrm{Gpc}$ box. Different colors represent various astrophysical variations. Observational data points are shown in blue for reference (see \cref{fig:clustering}). The luminosity threshold is \(L_{\mathrm{bol,thr}}=10^{45}\ \mathrm{erg\ s^{-1}}\). While the overall clustering patterns are broadly consistent across models, simulations with weak AGN feedback and with jet feedback exhibit slightly stronger clustering, particularly at $z = 1.0$.}
    \label{fig:phy_clustering}
\end{figure*}

In \cref{fig:phy_clustering}, we show the effects of different astrophysics variations on quasar clustering. Different colors represent different feedback models, and blue points are the observational data (same as \cref{fig:clustering}). The luminosity threshold is \(L_{\mathrm{bol,thr}}=10^{45}\ \mathrm{erg\ s^{-1}}\), which is different from \cref{fig:clustering} because of the small box size (so we can not compare them directly). 
We find that the clustering results are generally consistent across models, indicating that variations in parameters have relatively minor effects on the quasar clustering. However, at $z=1.0$, the simulations with jet feedback (Jet) and with weak AGN feedback (fgas+2\(\sigma
\)) exhibit slightly stronger clustering than the others. The origin of this difference is unclear and may be due to random fluctuations in the mass of the quasar hosts arising in different models. We note that at low $z$, the number of quasars drops, making these fluctuations in the clustering strength more likely.

\section{Shen et al. (2020) QLF fitting models}
\label{app:shenfits}

In this Appendix we summarize the observational QLF fitting methodology of \citet{shen2020}, which we use for constructing the cumulative quasar luminosity function (cQLF; see Sec. \ref{sec:qlf}). 

\citet{shen2020} parametrize the QLF with a double power-law function whose parameters are redshift-dependent:
\begin{equation}
\phi_{\mathrm{bol}}(L_{\rm bol};z)=\frac{dn(L_{\rm bol};z)}{d \mathrm{log_{10}} L_{\rm bol}}=\frac{\phi_{\star}(z)}{(L_{\rm bol}/L_{\star}(z))^{\gamma_1(z)}+(L_{\rm bol}/L_{\star}(z))^{\gamma_2(z)}}
\label{eq:QLF_obs}
\end{equation}
There are two global fitting approaches proposed by \citet{shen2020} to model this dependence. The first (``Global Fit A'') allows for a flexible, polynomial-like evolution of the faint-end slope, enabling it to vary dynamically with redshift and capture potential complexities at different epochs. In contrast, ``Global Fit B'' constrains the faint-end slope to evolve monotonically, following a power-law pattern that imposes a smoother and more consistent trend over time. In the first case, the parameters evolve with redshift as:

\begin{align}
& \gamma_1(z)=a_0 T_0(1+z)+a_1 T_1(1+z)+a_2 T_2(1+z) \\ 
& \gamma_2(z)=\frac{2 b_0}{\left(\frac{1+z}{1+z_{\mathrm {ref }}}\right)^{b_1}+\left(\frac{1+z}{1+z_{\mathrm {ref }}}\right)^{b_2}}
\end{align}
\begin{align}
& \log_{10} L_*(z)=\frac{2 c_0}{\left(\frac{1+z}{1+z_{\mathrm {ref }}}\right)^{c_1}+\left(\frac{1+z}{1+z_{\mathrm {ref }}}\right)^{c_2}} \\ 
& \log_{10} \phi_*(1+z)=d_0 T_0(1+z)+d_1 T_1(1+z) \\ 
& \left(T_0(x)=1, T_1(x)=x, T_2(x)=2 x^2-1\right)
\end{align}
In the second case, there is a different evolutionary model for the faint-end slope:
\begin{equation}
\gamma_1(z) = a_0\left(\frac{1+z}{1+z_\mathrm{red}}\right)^{a_1}
\end{equation}
The best-fit parameters of the two global evolution models can be found in \citet[][see their Table 4]{shen2020}.

In order to represent the cumulative uncertainty in the \citet{shen2020} observational model, we compute the expected cQLF by sampling the parameter space for each model $1000$ times. We assume that all parameters follow a Gaussian distribution, and we neglect any covariance between the parameters. We then take the 16th and 84th percentiles of the cQLF at each redshift. This gives two sets of constraints for the two global evolution models presented in \citet{shen2020}. 


\bsp	
\label{lastpage}
\end{document}